\documentclass[shortauth]{aa}
\usepackage{graphicx}
\usepackage{subcaption}
\usepackage{txfonts}
\usepackage[hidelinks]{hyperref}
\usepackage[normalem]{ulem}
\usepackage{orcidlink}
\usepackage{booktabs}
\usepackage{verbatim}
\usepackage{newtxtext,newtxmath}
\usepackage[T1]{fontenc}
\usepackage{graphicx}
\DeclareRobustCommand{\VAN}[3]{#2}
\let\VANthebibliography\thebibliography
\def\thebibliography{\DeclareRobustCommand{\VAN}[3]{##3}\VANthebibliography}

\title{Testing strengths, limitations, and biases of current pulsar timing arrays' detection analyses on realistic data}

\author{
Serena Valtolina \orcidlink{0009-0009-0322-2454} \inst{1}, Golam Shaifullah \orcidlink{0000-0002-8452-4834}\inst{2,3,4}, Anuradha Samajdar\orcidlink{0000-0002-0857-6018}\inst{5,6}, Alberto Sesana\orcidlink{0000-0003-4961-1606}\inst{2,3,7}}

\institute{
    Max Planck Institute for Gravitational Physics (Albert Einstein Institute), Leibniz Universit\"at Hannover, Callinstrasse 38, D-30167, Hannover, Germany\\
    \email{serena.valtolina@aei.mpg.de}
    \and
    Dipartimento di Fisica ``G. Occhialini", Universit{\'a} degli Studi di Milano-Bicocca, Piazza della Scienza 3, I-20126 Milano, Italy\label{unimib}
    \and
    INFN, Sezione di Milano-Bicocca, Piazza della Scienza 3, I-20126 Milano, Italy\label{infn-unimib}
    \and
    INAF - Osservatorio Astronomico di Cagliari, via della Scienza 5, 09047 Selargius (CA), Italy\label{inaf-oac}
    \and
    Institute for Gravitational and Subatomic Physics (GRASP), Utrecht University, Princetonplein 1, 3584 CC Utrecht, The Netherlands 
    \and
    Institut f\"{u}r Physik und Astronomie, Universit\"{a}t Potsdam, Haus 28, Karl-Liebknecht-Str. 24/25, 14476, Potsdam, Germany
    \and
    INAF - Osservatorio Astronomico di Brera, via Brera 20, I-20121 Milano, Italy\label{inaf-brera}
     }

\titlerunning{Testing strengths, limitations, and biases of current PTAs detection analyses on realistic data}
\authorrunning{S. Valtolina et al.}

\begin{document}
\label{firstpage}

\abstract{
State-of-the-art searches for gravitational waves (GWs) in pulsar timing array (PTA) datasets model the signal as an isotropic, Gaussian, and stationary process described by a power law. In practice, none of these properties are expected to hold for an incoherent superposition of GWs generated by a cosmic ensemble of supermassive black hole binaries (SMBHBs). This stochastic signal is usually referred to as the GW background (GWB) and is expected to be the primary signal in the PTA band.
We performed a systematic investigation of the performance of current search algorithms, using a simple power-law model to characterise GW signals in realistic datasets. We used, as the baseline dataset, synthetic realisations of timing residuals mimicking the European PTA (EPTA) second data release (DR2). Thus, we included in the dataset uneven time stamps, achromatic and chromatic red noise, and multi-frequency observations. We then injected timing residuals from an ideal isotropic, Gaussian, single power-law stochastic process and from a realistic population of SMBHBs, performing a methodical investigation of the recovered signal. 
We found that current search models are efficient at recovering the GW signal, but several biases can be identified due to the signal-template mismatch, which we identified via probability-probability (P\textendash P) plots and quantified using Kolmogorov-Smirnov (KS) statistics. 
We discuss our findings in light of the signal observed in the EPTA DR2 and corroborate its consistency with a SMBHB origin.}


\maketitle 

\section{Introduction}
\label{s:intro}
Recently, the pulsar timing array (PTA) community has announced the detection of a signal compatible with a gravitational wave (GW) origin. The latest datasets from the European PTA collaboration \citep[EPTA,][]{EPTA_III}, the Chinese PTA collaboration \citep[CPTA,][]{CPTA_new}, the North American Nanohertz Observatory for Gravitational Waves collaboration \citep[NANOGrav,][]{NANOGrav_new}, and the Parkes PTA collaboration \citep[PPTA,][]{PPTA_new} all show evidence of a red noise process that is common among pulsars and shows correlation properties expected for the long sought after stochastic nano-Hertz (nHz) GW background \citep[GWB,][]{HD}. 

Pulsar timing arrays are sensitive across the $10^{-9} - 10^{-7}$ Hz frequency range, where the dominant signal is expected to be an incoherent superposition of sinusoidal GWs emitted by a cosmic population of supermassive black hole binaries \citep[SMBHBs, ][]{Rajagopal_1995,jb2003,wl2003,Sesana_2008,Rosado_2015}. 
Nonetheless, a correlated low-frequency signal in PTA data can also arise from GWs generated by early Universe phenomena, such as the non-standard inflationary scenario breaking the slow-roll consistency relations \citep[e.g.][]{2007PhRvD..76f1302B,2008PhRvD..78d3531B,2011JCAP...06..003S}, cosmic string networks \citep[e.g.][]{2000PhRvL..85.3761D}, primordial curvature perturbations \citep[e.g.][]{1967PThPh..37..831T,1993PhRvD..47.1311M}, turbulence arising in the aftermath of quantum chromodynamics (QCD) phase transitions \citep[e.g.][]{1992PhRvD..45.4514K,2014PhRvL.112d1301H}, and cosmic domain walls \citep[e.g.][]{2014JCAP...02..031H}, or they might even originate from oscillations of the Galactic potential in the presence of ultra-light dark-matter \citep[ULDM,][]{Khmelnitsky_2014}. Those scenarios inspired numerous studies aiming to test new physics in the early Universe \citep[e.g.][]{2023JHEAp..39...81V,2023arXiv230614856M,2023arXiv230617022G,2023arXiv230617146K,2023arXiv230617147E,2023arXiv230617822C,2023arXiv230702399F,Franciolini2023pbf} and have also been scrutinised by the EPTA+InPTA and NANOGrav collaborations in two comprehensive interpretation articles \citep{EPTA_V,Afzal_2023}. 
early Universe interpretations sometimes stress that the detected signal has a higher amplitude and flatter slope than what is expected from an astrophysical population of SMBHBs. This is an observation that has been challenged by \cite{EPTA_V}, who showed that a signal with the detected properties can naturally arise from a realistic ensemble of SMBHBs. 

Under the hypothesis that the signal has an astrophysical, SMBHB origin, it is important to assess the performance of the state-of-the-art PTA GWB search and parameter estimation algorithms as implemented in the PTA analysis suite \texttt{enterprise} \citep{Ellis_2019} on realistic data. In fact, current pipelines are searching for a stochastic GWB described as a Gaussian, isotropic, and stationary process characterised by a power-law Fourier spectrum, imprinting a correlated red signal in PTA residuals. The peculiar feature that allows us to disentangle those GW-induced delays from other noises is the inter-pulsar spatial correlation, first derived by \citet{HD}. Although pipelines searching for individual continuous GWs (CGWs), anisotropy, and broken power-law spectra exist \citep{2012PhRvD..85d4034B,2012ApJ...756..175E,2015PhRvD..91h4055S,2020PhRvD.102h4039T}, the evidence for a GW-signal claimed by the different PTA collaborations is based on the aforementioned default assumptions. 
Conversely, none of these statistical properties are expected to hold for a signal produced by a cosmic ensemble of SMBHBs. Environmental effects and small number statistics are expected to produce spectra that significantly deviate from a smooth single power law \citep{Sesana_2008,2014MNRAS.442...56R}. Furthermore, strong individual CGWs might produce extra power at specific frequencies, also resulting in highly anisotropic and non-Gaussian signals \citep{2009MNRAS.394.2255S,2012ApJ...761...84R,2018MNRAS.477..964K}. Finally, the eccentricity of SMBHBs might break the assumption of stationarity \citep{2013CQGra..30v4014S}. 

In a nutshell, there is a clear signal-template mismatch situation, where single power-law, Gaussian, isotropic, stationary templates are used to search for a signal that is unlikely to satisfy any of these hypotheses. Assessing the influence of this basic fact on the reliability of the analysis outcome is of paramount importance,  especially in light of the recent PTA claims and of the numerous subsequent interpretation papers that rely on these results.

These issues were first investigated in \cite{Cornish_2016}, who demonstrated, using realistic GW models from \citep{Rosado_2015}, that the isotropy hypothesis has only a mild effect on the detectability of the GW signal. Here we take a decisive step forward by conducting a systematic investigation of signal recovery on mock data designed to capture all the complexity of real PTA datasets. We generate mock versions of the EPTA \texttt{DR2new} dataset, described in \cite{EPTA_I}, including uneven time-stamps, white noise, chromatic and achromatic red noise, and multi-frequency observations \citep{EPTA_II}. We then inject in our mock dataset two types of signals: (i) a Gaussian, isotropic, stationary GWB created with the \texttt{libstempo} \citep{libstempo}\footnote{\texttt{https://github.com/vallis/libstempo}} function \texttt{createGWB}, and (ii) an incoherent superposition of the residuals induced by a population of circular GW-driven SMBHBs. In this latter case, residuals from both the pulsar and Earth terms of each binary are added one by one to the data. This results in a total GWB Fourier spectrum which on average is well fitted by a power-law function of frequency, but is generally much more structured. In particular, bright sources close to Earth can produce very pronounced peaks in power, breaking signal Gaussianity and isotropy at higher frequencies \citep{Sesana_2008}. The rationale behind this choice is to test for differences in the performance of the GWB detection pipeline implemented in \texttt{enterprise} when (i) the injected signal matches the power-law template, and (ii) when there is a clear mismatch between the injected signal and power-law template.

The paper is structured as follows. In Section~\ref{s:methodsI} we describe the simulated dataset, from the intrinsic pulsar noise to the realistic GWB modelling pipeline, and we present the signal recovery model implemented in the \texttt{enterprise} package. Section~\ref{s:results} presents the results of our simulations, which include two sets of \texttt{createGWB} injections (a strong and a weak signal), and a set of injections from a realistic SMBHB population. In Section~\ref{s:specialcases}, we discuss in detail selected realisations of the realistic injections, which help in understanding biases and issues that can arise from the signal-template mismatch. As we were completing this work, \citet{Becsy_2023} presented an independent, parallel investigation that touches on several points examined here. We subsequently discuss similarities and differences between the two works, together with further developments and future directions in Section~\ref{s:conclusion}.

\section{Simulated data and analysis methods}
\label{s:methodsI}

The main goal of this work is to test our ability to recover a realistic GWB signal in a simulated data set with our existing model. 
To emulate the complexities of real data, we base all our simulations on the second data release from the EPTA collaboration \citep{EPTA_III,EPTA_I,EPTA_II}. In particular, we use the 25 best EPTA pulsars, selected following \citep{Speri_2022}. For these, we use the latest estimates of pulsar intrinsic red noise (RN) and dispersion measure (DM) variations to generate different simulated copies of the recently released \texttt{DR2new} dataset. \texttt{DR2new} is a reduced version of the entire second EPTA data release (\texttt{DR2full}), which includes only the last 10.3 years of observations, collected with the new generation wide-band backends. 

In the following, we describe how we simulate PTA data using \texttt{libstempo} tools, a python wrapper around the \texttt{TEMPO2} pulsar timing software \citep{Hobbs_2006,Edwards_2006}, and how we obtain realistic GW-induced residuals from state of the art population of SMBHBs.

\subsection{Timing model and pulsar noise properties}
\label{subsec:noise}
The timing models (TM) of the 25 pulsars are defined by the corresponding parameter files from the EPTA \texttt{DR2new}. We use \texttt{libstempo} to simulate times of arrival (ToAs) for a total observation time of 10.3 years, with a cadence defined by the observations, taken by the five European radio telescopes: the 100m Effelsberg radio telescope in Germany, the Lovell telescope at Jodrell Bank Observatory in the United Kingdom, the Nan\c{c}ay radio telescope operated by the Nan\c{c}ay Radio Observatory in France, the Sardinia Radio Telescope in Italy, and the Westerbork Synthesis Radio Telescope in the Netherlands. ToAs are simulated at two different frequency bands: 1400 and 2200 MHz. Having ToAs at different frequencies allows us to disentangle pulsar intrinsic RN from DM variations. We assume that an initial fit of the TM, obtained with \texttt{libstempo}, reduces it to a linear model where the coefficients are given by a design matrix. Following \cite{VanHaasteren_2009}, we analytically marginalise the likelihood over the TM parameter errors described by that linear model.

To make our simulations as close as possible to the real data, we include stochastic noise in our datasets. Stochastic noise in pulsar observations is customarily divided into three components: white noise, achromatic, and chromatic red noise. We include all three components in our simulations, following the analysis performed in \cite{EPTA_II}, based on the optimisation procedure outlined by \cite{Chalumeau_2022}. 

To model white noise, we distribute ToAs around the values predicted by the TM with a root-mean-square (rms) uncertainty given by: 
\begin{equation}
    \label{eq:efacequad}
    \sigma = \sqrt{{\rm EFAC}^2 \sigma_{\rm ToA}^2 + {\rm EQUAD}^2} \, ,
\end{equation}
where $\sigma_{\rm ToA}$ is the uncertainty value due to template-fitting errors obtained in EPTA DR2, the EFAC factor takes into account the ToA measurement errors. The EQUAD, added in quadrature, accounts for any other white noise, such as stochastic profile variations, and possible systematic errors. These parameters are specific for each observing backend. In our simulations, we defined EFAC and EQUAD parameters to be constant, setting EFAC = 1.0 and EQUAD = $10^{-6}$, and identical for all pulsars. The ToA uncertainties, $\sigma_{\rm ToA}$, used in our simulated data sets (see Table~\ref{tab:psrsnoise}) are the maximum likelihood estimates obtained in EPTA DR2 \citep[][]{chains_epta}.\\

The single-pulsar stochastic achromatic and chromatic red noises are time-correlated signals that can be modelled as a stationary Gaussian process with a power-law spectrum:
\begin{equation}
    \label{eq:RNdef}
    S_{\rm RN/DM}(f;A_{\rm RN/DM},\gamma_{\rm RN/DM}) = \frac{A_{\rm RN/DM}^2}{12\pi^2}\left( \frac{f}{yr^{-1}}\right)^{\gamma_{\rm RN/DM}} yr^3 \, ,
\end{equation}
where RN and DM refer to achromatic and chromatic red noise respectively.
The achromatic red noise does not depend on the observing radio frequency and is commonly used in single-pulsar noise models to characterise the long-term variability of the pulsar spin. Conversely, chromatic red noise depends on the observing radio frequency and is due to dispersion measure (DM) variations. In fact, during its propagation, the pulsar radio emission interacts with the ionised interstellar medium (IISM), the Solar System interplanetary medium and the Earth’s ionosphere. These interactions lead to frequency-dependent delays in the observed signal: 
\begin{equation}
    \label{eq:deltaDM}
    \Delta_{\rm DM} \propto \nu^{-2} {\rm DM} \, ,
\end{equation}
where $\nu$ is the radio observing frequency and DM is 
the path integral of the free-electron density along the line of sight to the pulsar. We take this effect into account in our timing model, which considers the DM value at a reference epoch together with its first and second derivatives. However, the inhomogeneous and turbulent nature of the IISM also induces stochastic variations in the DM value, which are modelled as chromatic red noise.  
DM variations become more and more important on the decade-long timescales of PTA data \citep[see e.g.][]{Keith_2012}.

We used the \texttt{libstempo} package to inject in each pulsar RN and DM with spectra given by Eq.~\eqref{eq:RNdef}. In particular, for each pulsar, we defined the values of $A_{\rm RN/DM}$ and $\gamma_{\rm RN/DM}$ to be equal to the recent maximum likelihood estimates from the EPTA DR2 \citep{EPTA_III}. Those values were obtained from 
joint analysis runs on the \texttt{DR2new} data set \citep{chains_epta} when an additional common red process was included alongside individual pulsar noise terms. We report a complete summary of the individual pulsar's noise parameter values in Table~\ref{tab:psrsnoise}.

As mentioned above, we simulate stochastic noise following the results of the customised noise model analysis carried out in \citet{EPTA_II}. Thus, for three pulsars J0030+0451, J1455-3330 and J2322+2057, we  simulate RN only. Five pulsars J0900-3144, J1012+5307, J1022+1001, J1713+0747, and J1909-3744 have both RN and DM variations, and for the remaining seventeen we include DM variations only.

\begin{table*}
\begin{center}
  \begin{tabular}{|c|c|c|c|c|c|c|c|c|}
  \hline
  Pulsar & \,$d$ [kpc] \, & \, cadence [days] \, & \, $\sigma_{\rm ToA}$ [$\mu s$] \, & \, noise model \,& \, ${\rm log}_{10}A_{\rm RN}$ \, & \, $\gamma_{\rm RN}$ \, & \, ${\rm log}_{10}A_{\rm DM}$ \, & \, $\gamma_{\rm DM}$ \,\\
  \hline
  \hline
  J0030+0451 & 1.1 & 1.1240 & 4.7321 & RN & -16.2802 & 0.3401 & - & - \\
  J0613-0200 & 1.2 & 2.1498 & 2.3190 & DM & - & - & -12.1099 & 2.2955 \\
  J0751+1807 & 0.9 & 1.5237 & 2.5138 & DM & - & - & -11.5769 & 2.1921 \\
  J0900-3144 & 1.3 & 0.6809 & 4.6006 & RN+DM & -12.7817 & 0.9483 & -11.8937 & 4.5107 \\
  J1012+5307 & 2.1 & 0.8983 & 3.1012 & RN+DM & -12.9283 & 1.4274 & -12.6575 & 3.9428 \\
  J1022+1001 & 0.3 & 2.0854 & 3.3792 & RN+DM & -17.3298 & 6.2477 & -11.4570 & 0.2273 \\
  J1024-0719 & 0.2 & 1.7813 & 4.5679 & DM & - & - & -11.9266 & 2.6198 \\
  J1455-3330 & 1.1 & 1.5463 & 11.6450 & RN & -13.1206 & 1.4706 & - & - \\
  J1600-3053 & 0.7 & 1.4481 & 0.8195 & DM & - & - & -12.8008 & 5.7666 \\
  J1640+2224 & 0.9 & 2.4334 & 5.7051 & DM & - & - & -11.4996 & 0.3000 \\
  J1713+0747 & 0.8 & 0.9403 & 0.9645 & RN+DM & -15.0142 & 3.0124 & -12.1243 & 1.6056 \\
  J1730-2304 & 1.3 & 3.2348 & 3.4269 & DM & - & - & -11.6328 & 1.5792 \\
  J1738+0333 & 1.5 & 5.0228 & 6.7555 & DM & - & - & -11.2123 & 1.5681 \\
  J1744-1134 & 2.2 & 2.4413 & 1.9787 & DM & - & - & -11.7996 & 0.7458 \\
  J1751-2857 & 1.5 & 12.3347 & 8.4155 & DM & - & - & -11.0296 & 1.0261 \\
  J1801-1417 & 0.6 & 9.7971 & 6.0490 & DM & - & - & -11.0496 & 2.2886 \\
  J1804-2717 & 1.1 & 5.8057 & 8.6381 & DM & - & - & -11.2871 & 0.0786 \\
  J1843-1113 & 1.3 & 5.1115 & 2.3503 & DM & - & - & -11.0364 & 2.3694 \\
  J1857+0943 & 0.6 & 3.2544 & 2.6596 & DM & - & - & -12.4013 & 4.6670 \\
  J1909-3744 & 1.3 & 1.6435 & 0.5931 & RN+DM & -16.8142 & 1.8981 & -11.9277 & 1.6375 \\
  J1910+1256 & 2.3 & 8.1784 & 4.1336 & DM & - & - & -11.9094 & 3.4703 \\
  J1911+1347 & 1.8 & 4.6388 & 2.0747 & DM & - & - & -12.1507 & 3.1562 \\
  J1918-0642 & 1.0 & 3.3059 & 2.8809 & DM & - & - & -12.3077 & 4.1191 \\
  J2124-3358 & 0.5 & 2.3498 & 6.6360 & DM & - & - & -11.4152 & 0.6247 \\
  J2322+2057 & 0.6 & 5.5817 & 13.8094 & RN & -15.1459 & 0.4594 & - & - \\
  \hline
  \end{tabular}
  \caption{Values for the distance, timing, and noise parameters of the 25 best EPTA pulsars. The final four columns list the maximum likelihood values obtained from the data set \texttt{DR2new}, using customised noise models when a common red noise is also included in the recovery model \citep{EPTA_III}. Thus, the missing RN and DM parameters refer to the fact that, according to the customised noise model, there is no relevant evidence for the given process in the EPTA \texttt{DR2new} data set.
  \label{tab:psrsnoise}}
\end{center}
\end{table*}

\subsection{GWB induced residuals: createGWB vs realistic SMBHB populations} 
\label{ss:realinjmodel}

Having addressed the system and pulsar-related properties of real data, the only thing necessary to complete our data set is the GW signal. Following \citet{Rosado_2015}, we generate the stochastic GWB-induced signal from the incoherent superposition of individual sinusoidal GW signals emitted by inspiralling SMBHBs. 

\subsubsection{Ideal signal with createGWB}
As described in \cite{phinney2001}, the characteristic strain of the GW signal produced by a population of circular, GW-driven SMBHBs is the integral of the energy emitted by each system over the differential number density of sources per unit redshift $z$, and chirp masses ${\cal M}$\footnote{The chirp mass is defined as ${\cal M} = (M_1M_2)^{3/5}/(M_1+M_2)^{1/5}$, where $M_2<M_1$ are the masses of the two black holes forming the binary. In the circular GW-driven approximation, at the quadrupolar order, the signal depends only on this combination of the two masses.}:
\begin{equation}
    \label{eq:hcgwb}
    h_{\rm c}^2 (f) = \frac{4}{\pi f^2}\int\int\int {\rm d}z \, {\rm d}M_1 \, {\rm d}q \,\frac{{\rm d}^2n}{{\rm d}z{\rm d}{\cal M}} \frac{1}{1+z} \frac{{\rm d}E_{gw}({\cal M})}{{\rm dln}f_r} \, ,
\end{equation}
where ${\rm d}E_{gw}({\cal M})/{\rm dln}f_r$ is the energy spectrum emitted by each source (binary). In the circular GW-driven approximation, we can rewrite the emitted energy spectrum as a function of the binary chirp mass and GW rest-frame frequency $f_r$:
\begin{equation}
    \label{eq:dE}
    \frac{{\rm d}E_{gw}({\cal M})}{{\rm dln}f_r} = \frac{\pi^{2/3}}{3}{\cal M}^{5/3}f_r^{2/3} \, .
\end{equation}
Here, $f_r$ is defined as $f_r = (1+z)f$ and is twice the binary Keplerian rest-frame frequency. By inserting Eq.~\eqref{eq:dE} into Eq.~\eqref{eq:hcgwb}, it is straightforward to show that 
\begin{equation}
    \label{eq:hcgwbpl}
    h_{\rm c}(f) = A_{\rm GWB} \left( \frac{f}{1yr^{-1}}\right)^{\alpha_{\rm GWB}} \, ,
\end{equation}
where $A_{\rm GWB}$ is a model-dependent amplitude of the signal at the reference frequency $f=1{\rm yr}^{-1}$ and $\alpha_{\rm GWB} = -2/3$. The corresponding spectral density $S_{\rm GWB}$ takes the form:
\begin{equation}
    \label{eq:gwbS}
    S_{\rm GWB} (f) = \frac{h_{\rm c}^2(f)}{12\pi^2f^3} = \frac{A_{\rm GWB}^2}{12\pi^2}\left( \frac{f}{1yr^{-1}}\right)^{-\gamma_{\rm GWB}} yr^3 \, ,
\end{equation}
with $\gamma_{\rm GWB} = 3 - 2\,\alpha_{\rm GWB} = 13/3$. We note that the form of Eq.~\eqref{eq:gwbS} is very similar to that of the intrinsic RN spectra (Eq.~\eqref{eq:RNdef}). In fact, in a PTA dataset, a stochastic GWB appears as a red noise that is common to all pulsars and induces a specific angular correlation among pulsar pairs. This correlation is expected to follow, on average, the \citet{HD} overlap reduction function.
Other examples of red signals correlated over all pulsars are clock and ephemeris errors, which produce, respectively, a monopole and a dipole correlation in the pulsar residuals \citep{tiburzi+16}.

Using \texttt{libstempo}, it is possible to inject a GWB signal into a PTA dataset through the function \emph{createGWB}. By default, this function simulates the GW-induced delays in a PTA dataset as a common RN, HD-correlated over pulsars and with a smooth power-law shaped spectrum with index $\alpha={13/3}$ (Eq.~\eqref{eq:gwbS}). The only input parameter supplied is the amplitude of the GWB signal. See discussion in \citet{Chamberlin_2015} for more details.

\subsubsection{Realistic signal from an SMBHB population}

Eq.~\eqref{eq:hcgwb} models the characteristic strain of the GW signal as a smooth continuous function of the frequency, as given by  Eq.~\eqref{eq:dE}. In reality, the expected astrophysical signal is the incoherent superposition of independent sinusoidal waves produced by an ensemble of SMBHB systems. For this case, Eq.~\eqref{eq:hcgwb} takes the form \citep{Sesana_2008}
\begin{equation}
 h_c^2(f) = \int_0^{\infty}{\rm d}z\int_0^{\infty}{\rm d {\cal M}}\frac{{\rm d}^3N}{{\rm d}z{\rm d}{\cal M}{\rm d}{\rm ln}f_r}h^2(f_r) \, ,
\label{eq:hch2_circ}
\end{equation}
where now ${\rm d}^3N/({\rm d}z{\rm d}{\cal M}{\rm d}{\rm ln}f_r)$ is the number of emitting systems per unit redshift, mass and logarithmic frequency interval, and the strain $h(f_r)$ is given by:
\begin{equation}
 h(f_r) =\sqrt{2(a^2+b^2)}\,\frac{(G{\cal M})^{5/3}(\pi f_r)^{2/3}}{c^4r} \, .
\label{eq:hrosado}
\end{equation}
Here, $r$ is the co-moving distance to the source and the functions $a=1+{\rm cos}^2\iota$ and $b=-2{\rm cos}\iota$ define the relative strength of the two strain polarisations as a function of the binary inclination angle $\iota$ \citep[see][for details]{Rosado_2015}.

Eq.~\eqref{eq:hch2_circ} can be further manipulated by considering that the signal comes from a finite collection of discrete sources, and the spectrum is practically constructed in discrete frequency bins $\Delta{f}=1/T$, where $T=10.3$yr is the duration of the PTA experiment. The characteristic strain can be thus written as
\begin{equation}
 h_c^2(f_i) = \sum_{j\in\Delta{f_i}}\frac{h_j^2(f_r)f_r}{\Delta{f_i}} \, ,
\label{eq:hch2_discrete}
\end{equation}
where $f_i$ is the central frequency of the bin $\Delta{f_i}$, and the sum runs over all the systems for which $f_r/(1+z) \in\Delta{f_i}$. 

To practically inject the signal from a cosmic population of MBHBs in our PTA timing residual, we proceed as follows.
The list of emitting binaries is randomly sampled from the numerical distribution ${\rm d}^3N/({\rm d}z{\rm d}{\cal M}{\rm d}{\rm ln}f_r)$, which is obtained from the empirical, observation-based models described in \cite{Sesana_2013}. The starting point is the galaxy merger rate, expressed as:
\begin{equation}
    \label{eq:galmergerrate}
    \frac{{\rm d}^3n_g}{{\rm d}z{\rm d}M_g{\rm d}q_g} = \frac{\phi(M_g,z)}{M_g {\rm ln}10}\,\frac{{\cal F}(z,M_g,q_g)}{\tau(z,M_g,q_g)} \, \frac{{\rm d}t_r}{{\rm d}z} \, ,
\end{equation}
where the subscript 'g' stands for 'galaxy'. Here $\phi(M_g,z)$ and ${\cal F}(z,M_g,q_g)$ are the galaxy mass function and the galaxy differential pair fraction function at redshift $z$. Those quantities can be directly measured from observations, while the typical merger time scale $\tau(z,M_g,q_g)$ can be inferred by detailed simulations of galaxy mergers. The galaxy mass is then related to the SMBH mass via scaling relations of the form:
\begin{equation}
    \label{eq:mgmbh}
    {\rm log}_{10}M_{\rm BH}=\alpha + \beta\, {\rm log}_{10} X \, ,
\end{equation}
where $X$ can be, depending on the model, the galaxy bulge mass, or its mid-infrared luminosity or velocity dispersion. We refer to \citet{Sesana_2013} for a list of those relations. Finally, SMBHs grow their mass through accretion in galaxy mergers. It is, however, unclear whether accretion mostly occurs before or after the SMBHB coalesces and, in the former case, whether accretion occurs preferentially on either of the two SMBHs.  

For this work, we model $\phi(M_g,z)$ from \cite{2013ApJ...777...18M}, ${\cal F}(z,M_g,q_g)$ from \cite{2009A&A...498..379D} and $\tau(z,M_g,q_g)$ from \cite{2008MNRAS.391.1489K}. The SMBH mass is related to the galaxy mass via the $M-\sigma$ relation given by \cite{2013ARA&A..51..511K} and we assume that accretion occurs on the two black holes prior to the final merger, with preferential accretion on the secondary hole \citep{2014ApJ...783..134F}. These choices result in a GWB with nominal amplitude $A_{\rm GWB}=2.4\times10^{-15}$ when computed with Eq. \eqref{eq:hcgwb} and expressed in the power-law form given by Eq. \eqref{eq:hcgwbpl}; this value is consistent with the one inferred from the EPTA \texttt{DR2new} analysis \citep{EPTA_III}. 

We use this observation-driven astrophysical model to numerically construct the function ${\rm d}^3N/({\rm d}z{\rm d}{\cal M}{\rm d}{\rm ln}f_r)$. We then draw 100 Monte Carlo samples of this function in the appropriate mass, redshift and frequency region of the parameter space. Each draw results in $\approx$100K binaries, which we refer to as a \emph{universe realisation}. For each binary, we specify its chirp mass, redshift, and GW-signal amplitude (Eq.~\eqref{eq:hrosado}) in the observer frame, sky location coordinates, inclination and polarisation angles. All binaries are assumed to be circular.

\begin{figure}
	\includegraphics[width=\columnwidth]{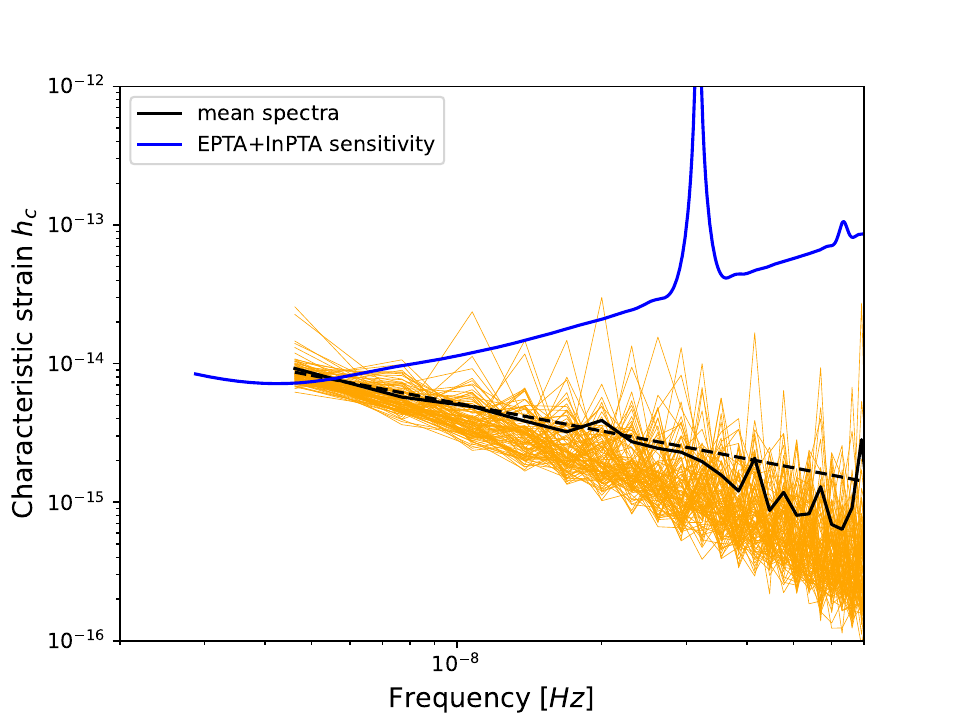}
    \caption{Characteristic strain as a function of frequency for 100 different realisations of a GWB with nominal amplitude $2.4 \times 10^{-15}$. Each orange line corresponds to a different realisation (SMBHB population). The solid black line is the mean of those realisations, while the dashed black line highlights the nominal $f^{-2/3}$ spectrum. The sensitivity curve is derived in \citet{3Ppaper}.}
    \label{fig:hc_all}
\end{figure}

Once the population of SMBHBs is defined, we construct the overall GW signal by directly injecting in the time domain the deterministic residulas imprinted by each individual system in the PTA dataset. To this end, we developed a custom \emph{injection} pipeline, written partly in \texttt{python} (using  \texttt{libstempo} functions) and partly in \texttt{fortran}. The script allows the user to take idealised pulsar timing models and produce ToAs for a given observing time span, add to each pulsar specific source noise as described in Section~\ref{subsec:noise}, and then add excess delays due to each of the SMBHB in the specified population. We inject both pulsar and Earth terms in the residuals, following the prescription of \cite{2016MNRAS.455.1665B}. 
Similar injection pipelines have been applied to NANOGrav-like datasets in \citet{Becsy_2022} and \citet{Becsy_2023}.

We can visually verify how this GWB signal definition method affects the spectra. The obtained characteristic strain amplitude for each of the 100 mock realisations of universe described above can be computed in the frequency domain from Eq.~\eqref{eq:hch2_discrete}, and is shown in Fig.~\ref{fig:hc_all}. As expected, the signal is much more structured than a plain $f^{-2/3}$ nominal power law computed through Eq.~\eqref{eq:hcgwb}, with prominent spikes associated with rare, massive and (or) nearby sources. We note that the square-averaged signal sits on the theoretical curve, but there is considerable variance among different universe realisations.

\subsection{Recovery model and analysis methods}
\label{sub:recovery}

To test the performance of current PTA GWB search and parameter estimation pipelines, we perform a set of inference analyses by using Bayes' theorem on the realistic datasets defined in Section \ref{s:methodsI}. We search for individual and common noise parameters estimating model parameters characterised by their posterior probability distribution functions (PDFs), including those of RN, DM, and common noise. We then carry out Bayes-factor (BF) evaluations between HD-correlated and common, uncorrelated RN (CURN) models; and reconstruct the angular correlation from the data.

We conducted an in-depth analysis of each dataset using \texttt{enterprise} \citep[\emph{Enhanced Numerical Toolbox Enabling a Robust PulsaR Inference SuitE, }][]{Ellis_2019}, a pulsar timing analysis package including functionalities for timing model evaluation, pulsar noise analysis, and GW searches. For each simulated array of pulsars we defined a noise model including: (i) EFAC and EQUAD parameters fixed respectively at 1.0 and 1e-6 for all pulsars, (ii) RN and DM variations according to the customised noise model used in EPTA \texttt{DR2new} (see Table~\ref{tab:psrsnoise}), (iii) a common red noise process. We set the number of coefficients (i.e. the number of modes in the Fourier domain) for modelling the RN and DM processes in each pulsar to be, respectively, 30 and 100. We modelled DM variations as a Gaussian process.

Using this recovery model, we carried out Bayesian inference of the model parameter space using the Markov Chain Monte Carlo (MCMC) sampler included in \texttt{enterprise}: \texttt{PTMCMCSampler} \citep{Ellis_2017}.  
Since we fixed the white noise parameters for each pulsar, the total number of parameters for the sampling is 62: 60 from pulsar intrinsic noise parameters and two parameters, log amplitude and slope, for the common red noise. 

To decrease computational costs, we employed the \emph{reweighting} method introduced in \citet{Hourihane_2023}. We first computed approximate posteriors defining the common process as an uncorrelated red noise. Thus, by temporarily ignoring the cross-correlation terms, the covariance matrix becomes block-diagonal, resulting in a much faster sampling. We then reweighted the obtained chains of samples to get the exact posteriors (corresponding to an HD-correlated common red process) via importance sampling. Besides being much faster than a direct search for an HD-correlated common signal, this method also provides an accurate estimate of the Bayes factor between the HD and the CURN models. In fact, the average of the weights computed for all samples is equal to the ratio of the marginal likelihood (or evidence) of the two considered models. While this method is mathematically exact, there are some limitations when the likelihood changes significantly between the two models. For example, as the GWB amplitude increases, the weights distribution becomes broader and the sampling efficiency of the method decreases. \citet{Hourihane_2023} present a very detailed study of the limits of this method and conclude that the Bayes factor estimate stays robust up to $BF > 10^6$. See \citet{Hourihane_2023} for a more detailed description.

To test the limits imposed by the sensitivity of the dataset and the consequences in the inference analysis, for some realisations we repeated the parameter estimation runs considering different numbers of Fourier components for the common process (see Section~\ref{ss:case2e3} for more details). The results presented in Sec.~\ref{ss:highSNR} and~\ref{ss:createvsrealinj} are obtained considering only the first nine frequency bins while searching for the common process.

We use uniform priors for the slope parameters and log-uniform priors for the amplitudes of the noise components. The prior ranges for the intrinsic noise and common process parameters are listed in Table~\ref{tab:priors}.

\begin{table}
\begin{center}
  \begin{tabular}{c|c|c}
  \hline
  Parameter & Prior type & Range\\
  \hline
  \hline
  $\gamma_{\rm RN}$, $\gamma_{\rm DM}$, $\gamma_{\rm CURN}$  & Uniform & $[ 0, \, 7 ]$\\
  $A_{\rm RN}$ & log-Uniform & $[10^{-18}, \, 10^{-11}]$ \\
  $A_{\rm DM}$ & log-Uniform & $[10^{-18}, \, 10^{-8}]$ \\
  $A_{\rm CURN}$ & log-Uniform & $[10^{-15.5}, \, 10^{-13.5}]$ \\
  \hline
  \end{tabular}
  \caption{Prior distributions for the Bayesian inference analyses.\label{tab:priors}}
\end{center}
\end{table}

For each simulated PTA dataset, we also computed the induced angular correlation in the timing residuals between pulsar pairs. We followed the method used for the frequentist analysis in EPTA \texttt{DR2new} \citep{EPTA_III}. We used the \emph{Optimal Statistic} (OS) framework developed by \citet{Anholm_2009}, \citet{Demorest_2013}, and \citet{Chamberlin_2015}, with the noise marginalisation described in \citet{Vigeland_2018}. We also computed the mean correlation and variance in the correlation recovery between different realisations of the same GWB signal, and compared the results with the theoretical predictions from \citet{Allen_2023}. We followed the prescriptions in \citet{AllenRomano} when computing the average over pulsar pairs.

\section{Results}
\label{s:results}

Using the framework described in the previous section, we generate three sets of 100 mock EPTA \texttt{DR2new} datasets, for a total of 300 simulations. In each of these 300 simulations, the individual pulsar DM and RN are generated as a random realisation of a stochastic process described by the power-law spectra of Eq.~\eqref{eq:RNdef}, with amplitude and slope fixed to the ML value of the customised noise analysis performed in \cite{EPTA_II} as reported in Table \ref{tab:psrsnoise}. The three sets of simulations differ for the injected GW signal:
\begin{itemize}
\item \emph{LoudGWB\_set.} We use  \emph{createGWB} to inject a loud, stochastic GWB with $A_{\rm GWB}=5\times 10^{-15}$. This signal is easily detectable in the \texttt{DR2new} dataset and serves as a benchmark to test our simulations and analysis pipeline.
\item \emph{CreateGWB\_set.} We use again \emph{createGWB} to inject a stochastic GWB with $A_{\rm GWB}=2.4\times 10^{-15}$, consistent with the signal observed in \cite{EPTA_III}. These simulations are meant to test the pipeline in the regime of a relatively weak signal that matches the template used in the likelihood evaluation.
\item \emph{SMBHB\_set.} We inject individual residuals from an astrophysically motivated SMBHB population producing a GWB with a nominal average amplitude of $A_{\rm GWB}=2.4\times 10^{-15}$. In this case, however, the signal is very different from the template used in the analysis pipeline, allowing us to investigate limitations and biases due to a mismatch between the signal present in the data and the model used in the analysis. 
\end{itemize}

In Section~\ref{ss:highSNR} we discuss the performance of the pipeline applied to the \emph{LoudGWB\_set}; we then move to the comparison between the \emph{createGWB\_set} and the \emph{SMBHB\_set} in Section~\ref{ss:createvsrealinj}.

\subsection{The LoudGWB\_set of simulations: A benchmark for recovery convergence}
\label{ss:highSNR}

\begin{figure}
	\includegraphics[width=\columnwidth]{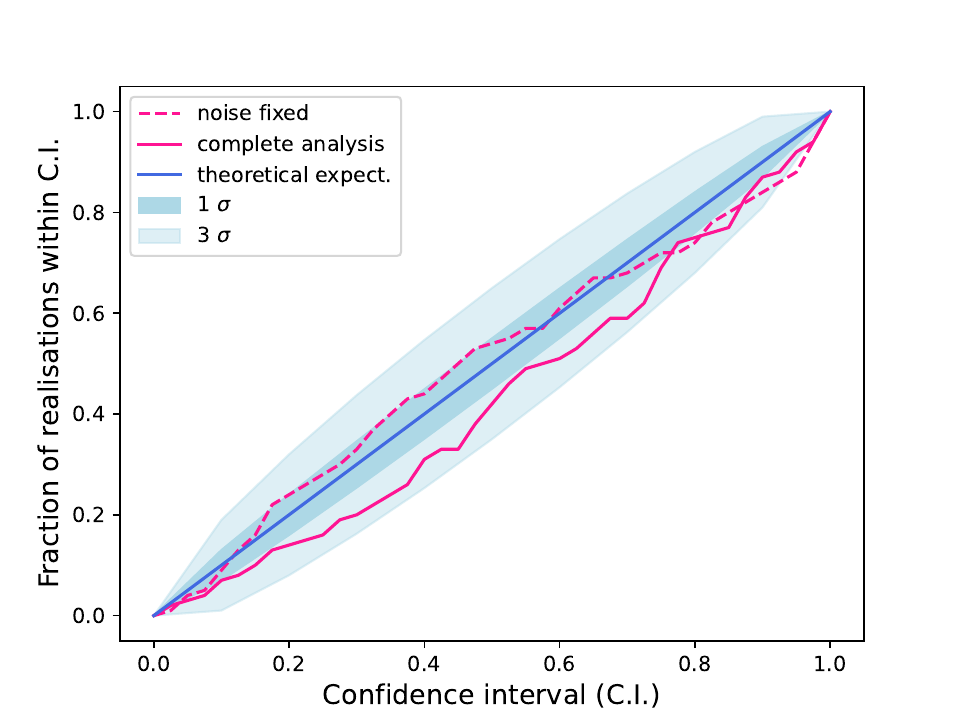}
    \caption{P\textendash P plot for the GWB-amplitude recovery for the {\it LoudGWB\_set} of simulations. The dashed line refers to the inference runs where all the intrinsic noise parameters are fixed in a noise dictionary and the amplitude of the GWB is the only free parameter. The solid line is the result of MCMC runs over all 62 noise parameters. The theoretical expectation for an unbiased recovery and the predicted variance of one- and three- $\sigma$ are represented in different shades of blue.}
    \label{fig:PP_highSNR}
\end{figure}
\begin{figure}
	\includegraphics[width=\columnwidth]{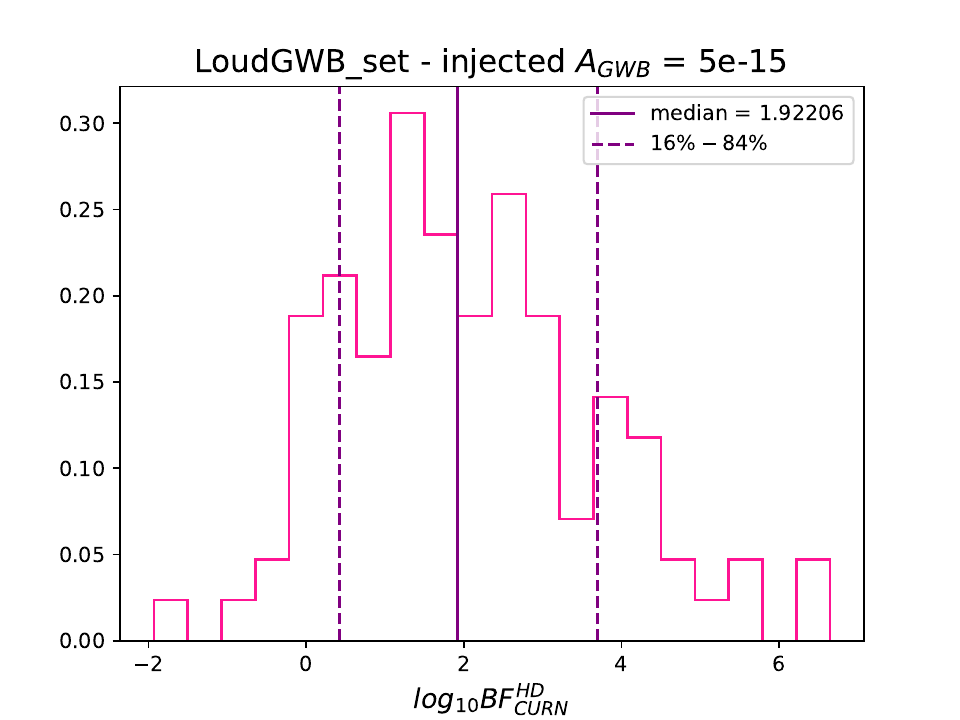}
    \caption{${\rm log}_{10}{\rm BF}$ distribution of HD vs CURN for the 100 realisations of the {\it LoudGWB\_set}. The vertical lines show the median (solid) and the 16th and 84th percentile (dashed) of the distribution.}
    \label{fig:bf_highA}
\end{figure}
\begin{figure}
	\includegraphics[width=\columnwidth]{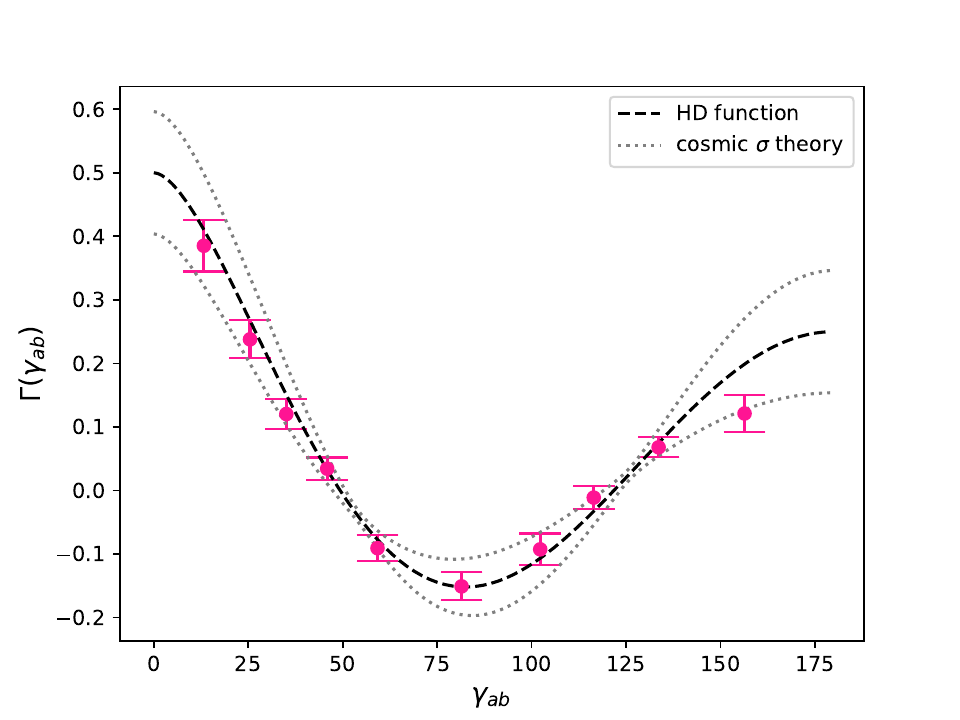}
    \caption{HD correlation recovery for the {\it LoudGWB\_set}. Each data point corresponds to the average of the optimal correlation estimators in that bin over all realisations. The expected cosmic variance is derived in \citet{Allen_2023}.}
    \label{fig:HD_highA}
\end{figure}

As already mentioned, this set of simulations aims to define a benchmark for recovery convergence. In fact, the injected high GWB amplitude ($A_{\rm GWB}=5\times 10^{-15}$) is expected to be relatively easy to recover and disentangle from pulsar noise. 

Some useful tools to validate the performance of our model
are the cumulative distributions of the number of times the injected value lies within a credible interval, the so-called probability-probability (P\textendash P) plots \citep[see][and others]{ppplots2006,talts2020validating,wilkgnanadesikan68}. 
These plots show on the \emph{y} axis the fraction of times in which the nominal value (in our case, the injected $A_{\rm GWB}$, normalised at the frequency of $1/1yr$) lies within the credible interval indicated on the \emph{x} axis.
In the case of unbiased inference, the data points follow the diagonal of the plot parameter space. 

A P\textendash P plot for the inferred amplitude of the recovered GWB is shown in Fig.~\ref{fig:PP_highSNR}.
The theoretical expectation and the predicted variance (one-$\sigma$ and three-$\sigma$ levels) are shown in different shades of blue. The dashed line is obtained from inference runs where all the pulsar's intrinsic noise parameters and the slope of the GWB signal are fixed to the nominal injected values (Table~\ref{tab:psrsnoise}, $\gamma_{\rm GWB} = 13/3$). Thus, the only free parameter is the amplitude of the GWB signal. In this case, the obtained distribution follows the diagonal within the one-$\sigma$ variance and there is no evident bias in the recovery of $A_{\rm GWB}$. This is expected since the GWB signal is injected via the {\it createGWB} function, which simulates a background with a spectrum very close to the nominal power law. The solid line, instead, is obtained from the
full analysis sampling of all the 62 model parameters (the pulsars intrinsic RN and DM variation parameters, see Table~\ref{tab:psrsnoise}, and the two GWB parameters). Although the recovered distribution is always within the three-$\sigma$ expected interval, it systematically lies below the diagonal. This may indicate a slightly biased recovery towards lower amplitudes (and, consequently, higher $\gamma_{\rm GWB}$) for the GWB spectrum. Although not particularly worrying, the origin of this potential bias is unknown and it might be due to leakage of power across different noise components.

Thanks to the reweighting method, we can also show the distribution of the obtained ${\rm log}_{10}{\rm BF}$ for the HD-correlated model versus CURN. The results are shown in Fig.~\ref{fig:bf_highA}. The median of the distribution (vertical solid line) is at  ${\rm log}_{10}{\rm BF}=1.91$, which corresponds to a BF$\sim 81$. The spread of the distribution highlights the impact of the stochastic nature of the pulsar noise in the signal recovery. The central 68\% of the BF distribution spans more than three orders of magnitude, and depending on the specific realisation of the noise, the data can either provide decisive evidence of a GWB or an inconclusive result.

Finally, we compute the angular correlation induced in the residuals of each pulsar pair for all realisations. For an array of 25 pulsars, there are 300 independent pairs. For each realisation, we define ten angular separation bins of 30 pulsar pairs each, and compute the mean and variance of the correlation. The calculation follows the prescriptions in \citet{AllenRomano}, taking into account the covariance between different pulsar pairs in the same angular separation bin. We then compute the mean over the whole \emph{LoudGWB\_set} as the average, in each angular separation bin, of such optimised correlation estimators. We present the result in Fig.~\ref{fig:HD_highA}, where each data point is computed over 3000 pulsar pairs: 30 for each realisation. For comparison, we also show the cosmic variance limit derived in \citet{Allen_2023} (see their Eq.~4.8 and~G11). The pulsar variance contribution to the expected variance of the HD recovery is minimised by the weighted-average method described in \citet{AllenRomano}; thus, in this case the cosmic contribution is the only significant one to compare our results with.

We note that the mean correlation estimated in each bin typically lies very close to the expected HD correlation. This is expected for such a loud GWB signal ($5\times 10^{-15}$). The only exception is the very last bin, which is difficult to constrain due to the limited number of pulsar pairs available at wide angular separation, forcing an averaging procedure over a wide bin.

\subsection{Ideal vs real: Comparing the createGWB\_set and the SMBHB\_set}
\label{ss:createvsrealinj}

Having assessed the performance of the analysis pipeline on a loud, ideal signal, we now turn to the comparison of the signal recovery for the  {\it createGWB\_set} and the {\it SMBHB\_set}.

\begin{figure}
    \includegraphics[width=0.53\textwidth]{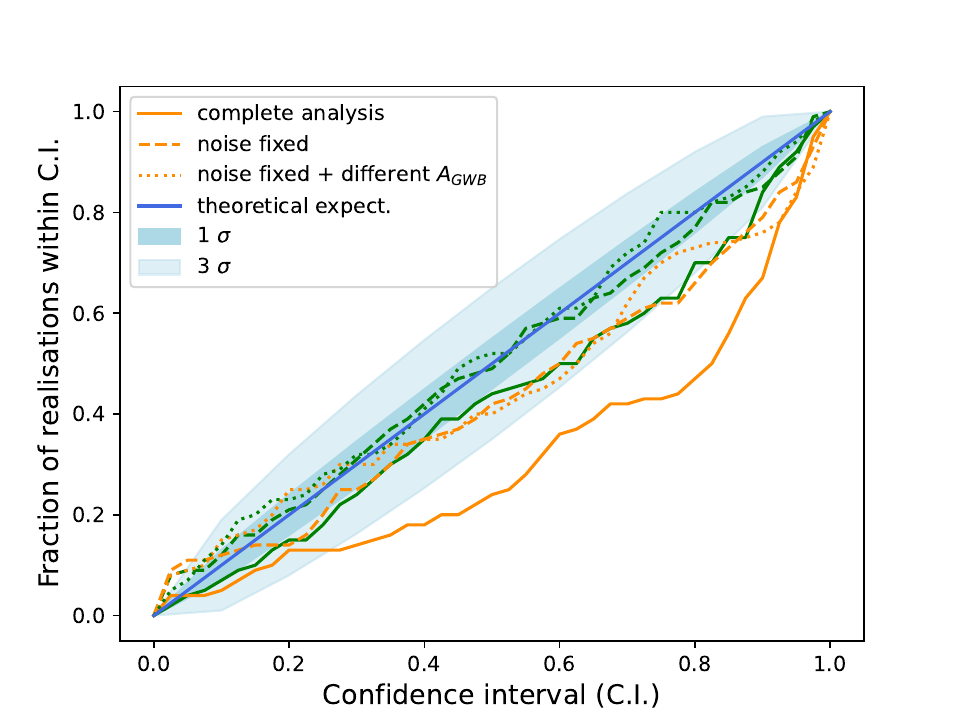}
    \caption{P\textendash P plot for the recovery of $A_{\rm GWB}$ from the {\it SMHBH\_set} (orange) and {\it createGWB\_set} (green). The solid lines are obtained from the ${\rm log}_{10}A_{\rm GWB}$ posteriors of the MCMC runs over all 62 noise parameters of the pulsars array. The dashed lines are obtained by fixing all noise parameters in the recovery model; thus, $A_{\rm GWB}$ is the only free parameter. The dotted lines are also from posteriors obtained sampling only over $A_{\rm GWB}$, but in a dataset where the realisations do \emph{not} all have the same spectral amplitude. Here, the injected background has an amplitude value extracted from the prior distribution used in the recovery.}
    \label{fig:mainPP}
\end{figure}

\begin{table}
\begin{center}
  \begin{tabular}{|c|c|c|c|}
  \hline
  Model 1 & \,Model 2 \, & \, KS stat. \, & \, p-value\,\\
  \hline
  \hline
  {\it SMBHB\_set} & diagonal & 0.1707 & 0.5945\\
  noise fixed & &  & \\
  \hline
  {\it createGWB\_set} & diagonal & 0.0732 & 0.9999\\
  noise fixed & &  & \\
  \hline
  {\it SMBHB\_set} & diagonal & 0.3590 & 0.0125\\
  full analysis & & & \\
  \hline
  {\it createGWB\_set} & diagonal & 0.1538 & 0.7523\\
  full analysis & &  & \\
  \hline
  {\it SMBHB\_set} & {\it createGWB\_set} & 0.1463 & 0.7789\\
  noise fixed & noise fixed &  & \\
  \hline
  {\it SMBHB\_set} & {\it createGWB\_set} & 0.2821 & 0.0897\\
  full analysis & full analysis & & \\
  \hline
  \end{tabular}
  \caption{KS statistic and correspondent p-values obtained from comparisons between different distributions shown in Fig.~\ref{fig:mainPP}. With {\it diagonal} we refer to the theoretical prediction for unbiased recoveries, which corresponds to the diagonal of the plot. The other distributions analysed are the dashed and solid lines of that P\textendash P plot. The p-value in the second raw is almost perfect ($0.9999$); this confirms the good fit between the $A_{\rm GWB}$ recovery for the {\it createGWB\_set} when the other noise parameters are fixed. \label{tab:ks}}
\end{center}
\end{table}

\begin{figure*}
    \centering
	\includegraphics[width=\columnwidth]{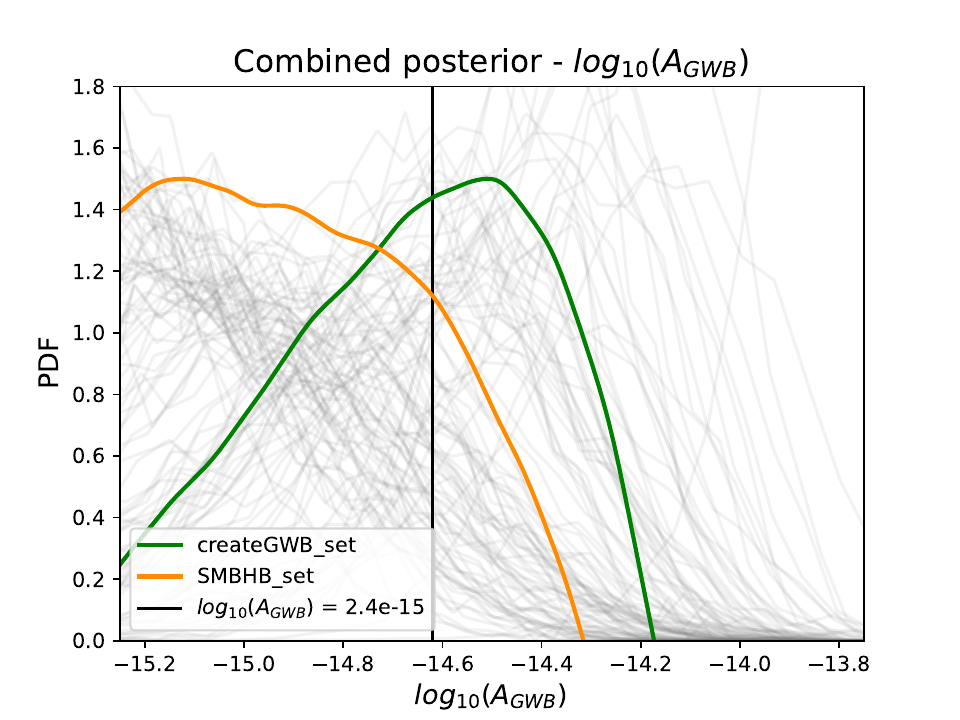}
	\includegraphics[width=\columnwidth]{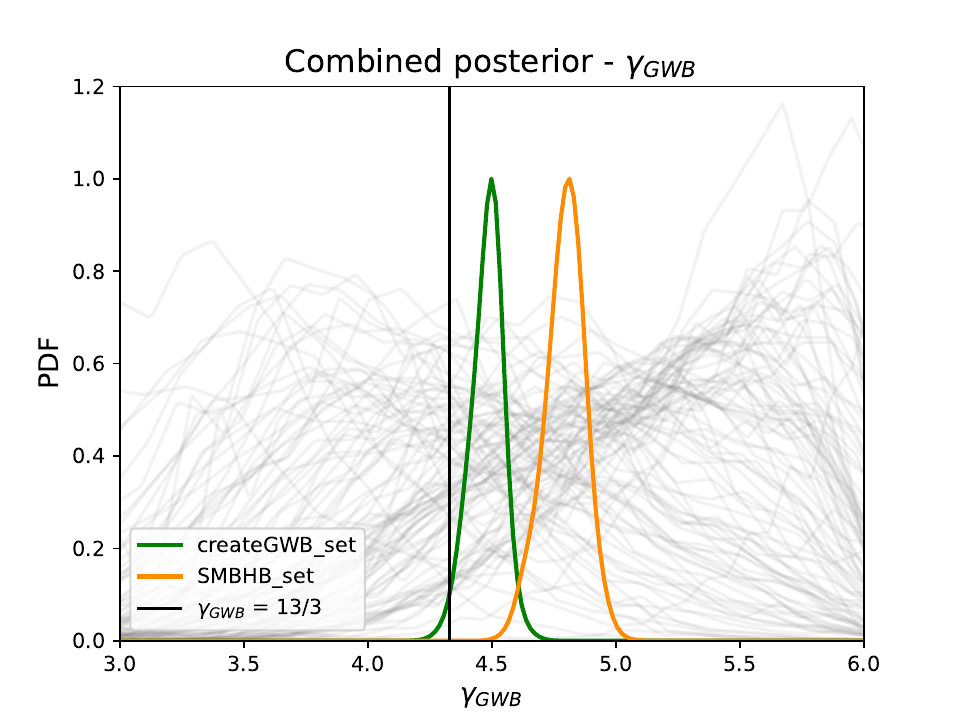} 

    \caption{Combined posterior for ${\rm log}_{10}A_{\rm GWB}$ (left panel) and $\gamma_{\rm GWB}$ (right panel) parameter for the {\it createGWB\_set} (green) and the {\it SMBHB\_set} (orange). The injected values, marked by the black vertical lines, are  $A_{\rm GWB} = 2.4\times10^{-15}$ and $\gamma_{\rm GWB} = 13/3$. The grey posteriors in the background are the ones obtained from the individual realisations of the {\it SMBHB\_set}. }
    \label{fig:combinedAG}
\end{figure*}

\begin{figure}
	\includegraphics[width=\columnwidth]{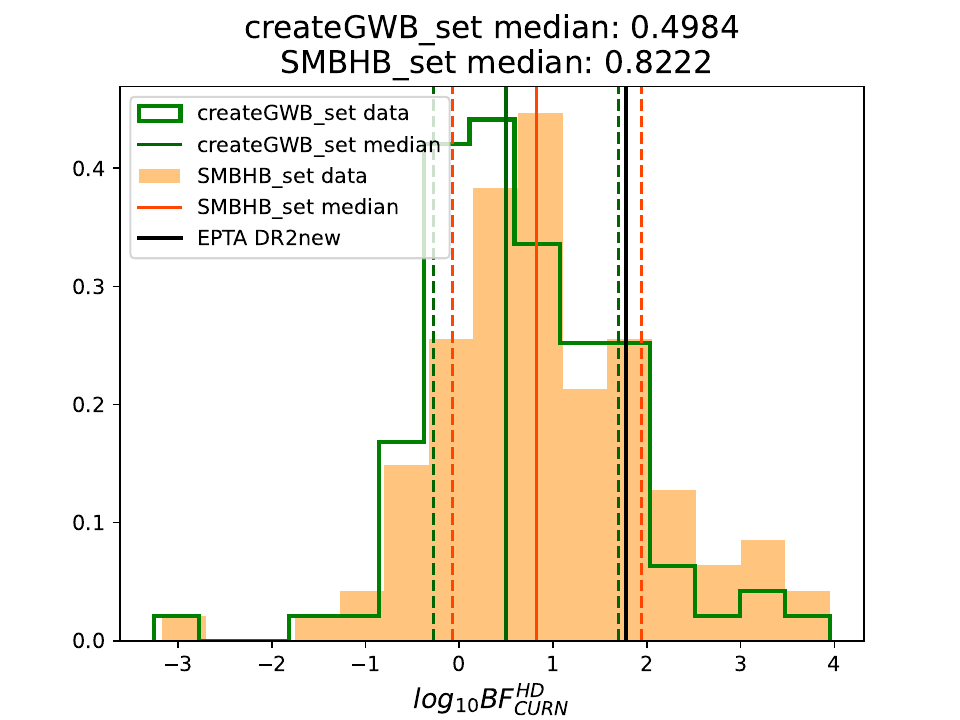}
    \caption{Histograms of the ${\rm log}_{10}{\rm BF}_{\rm CURN}^{\rm HD}$ obtained from the reweighting analysis on the {\it createGWB} (green) and  {\it SMBHB\_set} (orange). The solid vertical lines show the medians of the distributions, while the dashed ones refer to the 16th and 84th percentiles. The black vertical line corresponds to the latest EPTA estimate for \texttt{DR2new}: ${\rm BF}_{\rm CURN}^{\rm HD} \sim 60$.}
    \label{fig:logBF_2.4}
\end{figure}

\begin{figure*}
    \centering
	\includegraphics[width=\columnwidth]{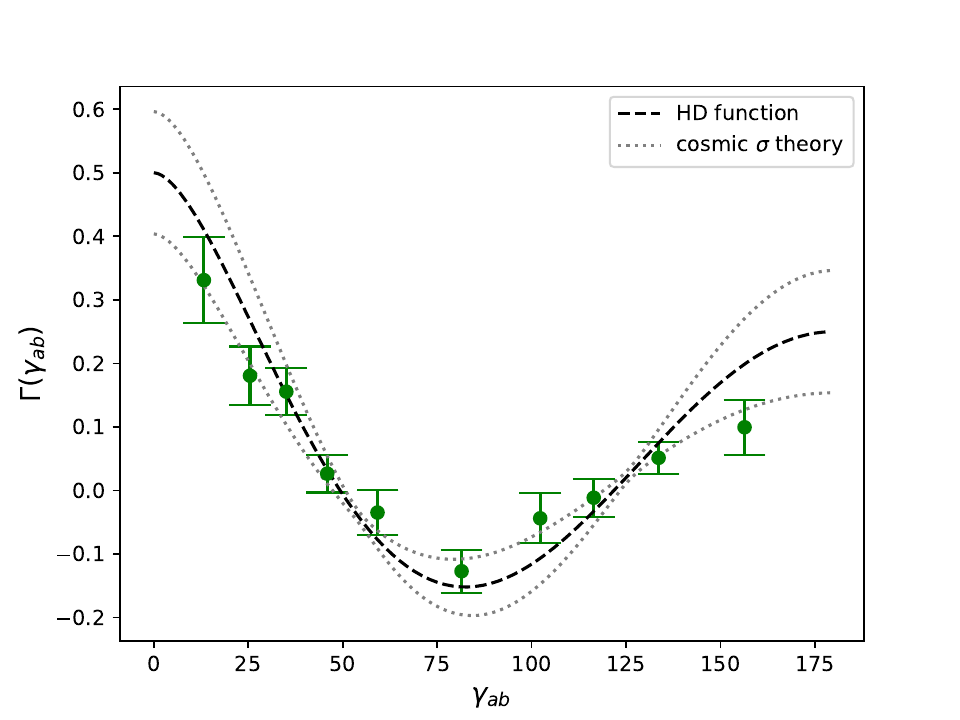}
	\includegraphics[width=\columnwidth]{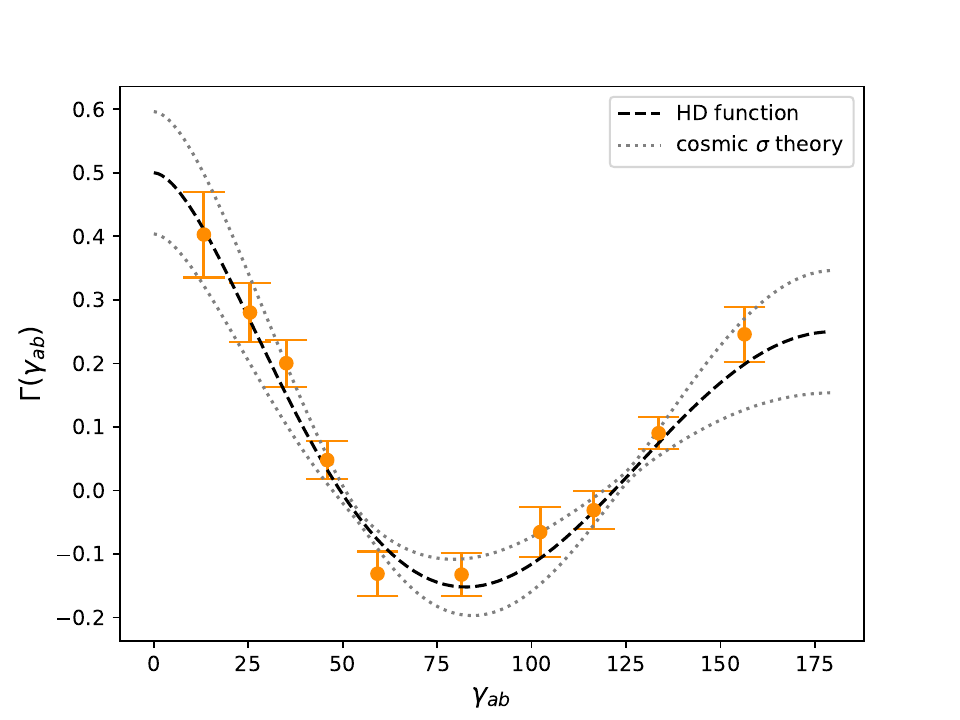}
    \caption{Mean and variance of the angular correlation in the timing residuals of the {\it createGWB\_set} (left panel) and for the {\it SMBHB\_set} (right panel). Each data point corresponds to the average of the optimal correlation estimators in that bin over all realisations. We also show the expected cosmic variance derived in \citet{Allen_2023}.}
    \label{fig:HD_createSMBHB}
\end{figure*}

As in the previous section, we start by constructing the P\textendash P plot for the $A_{\rm GWB}$ parameter, shown in Fig.~\ref{fig:mainPP}. For both data sets, the nominal value of $A_{\rm GWB}$ for each realisation is $2.4\times10^{-15}$, at the reference frequency of $1/1yr$. In the {\it SMBHB\_set} this value corresponds to computing the GWB through Eq. \eqref{eq:hcgwb} and expressing it in the power-law form given by Eq. \eqref{eq:hcgwbpl}.
The green lines are for the {\it createGWB\_set}, while the orange ones are for the {\it SMBHB\_set}. As in Fig.~\ref{fig:PP_highSNR}, the solid lines are computed using the posterior distributions of the MCMC runs sampling over all the 62 noise parameters of the array, while the dashed lines are obtained by searching over $A_{\rm GWB}$ keeping all other noise parameters fixed to the nominal value. 
When fixing all noise parameters, the {\it createGWB\_set} closely follows the diagonal, indicating an unbiased recovery of the signal amplitude. Conversely, the {\it SMBHB\_set} tends to be consistently below the diagonal, also crossing the three-$\sigma$ confidence interval in the upper-right corner.
When sampling over all the noise parameters, this bias is enhanced. As for the {\it LoudGWB\_set}, the distribution for the {\it createGWB\_set} (solid green) is still within the three-$\sigma$ confidence interval, although systematically below the diagonal. The situation gets more extreme for the {\it SMBHB\_set} (solid orange), which is dramatically biased towards the lower amplitude of the GWB spectra. 

To quantify the distances between the different distributions shown in Fig.~\ref{fig:mainPP}, we used the \texttt{scipy} package \texttt{scipy.stats.kstest} \citep{scipy} to perform a series of non-parametric Kolmogorov-Smirnov tests (KS test). The test returns the maximum difference between two distributions and an estimate of the p-value under the null hypothesis that the two distributions are identical. Results are summarised in Table~\ref{tab:ks} and highlight the inconsistency of the signal recovery in the  {\it SMBHB\_set}. 

We note that in these analyses there is a difference between the prior defined in the recovery model, ${\rm log}_{10}A_{\rm GWB}$ uniform in $[-15.5, -13.5]$, and the one from which the injected values are selected (basically a delta function centred in $A_{\rm GWB} = 2.4\times10^{-15}$). 
Since the latter prior is much narrower and completely included in the former, the statistical significance of the P\textendash P plots are unaffected.
To test this statement, we generated a new set of simulations where the pulsar's intrinsic noises are the same as for the other datasets and fixed, but now the GWB signal amplitudes have values which are randomly chosen from the uniform prior ${\rm log}_{10}A_{\rm GWB}$: $[-15.5, -13.5]$. The results are shown in Fig.~\ref{fig:mainPP} as dotted lines (green for the {\it createGWB\_set}, orange for the {\it SMBHB\_set}). As expected, there is no significant discrepancy between those lines and the dashed ones (same analysis on data where the GWB signal is the same in all realisations).

Since we have 100 simulations, each with a different realisation of population parameters but the same injected parameters of the GWB signal, we can utilise Bayes' theorem to obtain the combined and better constrained posterior PDFs for ${\rm log}_{10}A_{\rm GWB}$ and $\gamma_{\rm GWB}$. 
In Fig.~\ref{fig:combinedAG} we show the combined posteriors for ${\rm log}_{10}A_{\rm GWB}$ and $\gamma_{\rm GWB}$ for both the {\it createGWB\_set} (green) and the {\it SMBHB\_set} (orange). 
The posterior for the slope parameter is better constrained than that for the amplitude. While both parameters are compatible with the nominal injected value within $\sim$ one-$\sigma$ for the {\it createGWB\_set}, this is not the case for the {\it SMBHB\_set}. 
The orange posteriors are clearly biased towards steeper spectra with lower amplitudes, consistent with what is shown by the P\textendash P plots. 

Systematic biases in the signal recovery for the {\it SMBHB\_set} can be traced back to the $h_c$ distribution of the individual realisations shown by the orange lines in Fig.~\ref{fig:hc_all}. These realisations lie preferentially below the expected $f^{-2/3}$ line, featuring a steeper spectrum. Only sporadically, loud individual sources result in excess power at specific frequencies. When this happens, a power-law fit to the data can result in a flatter spectrum, above the $f^{-2/3}$ line. This is a general feature of realistic SMBHB populations characterised by a sparse, high-mass tail of loud GW sources. Although most signal realisations show a deficiency of power at high frequency, the few which feature loud sources ensure that the average signal amplitude sits on the expected $f^{-2/3}$ power law. It follows that the typical realisation of a nominal $f^{-2/3}$ power-law signal has in reality, a steeper spectrum. There is, therefore, a mismatch between the theoretically smooth GWB signal used in the recovery model and the real signal produced by a discrete ensemble of SMBHBs, which can lead to systematic biases in the signal recovery and erroneous interpretation of the results. 

We note that, during our whole analysis, we fixed the reference frequency for the recovery of the GWB amplitude to $1/1yr$. Changing this frequency to a lower one would result in a weaker dependence of $A_{\rm GWB}$ upon $\gamma_{\rm GWB}$, but the one-dimensional posterior of the slope parameter remains unchanged. 
Reanalysing our data sets with the reference frequency set at $1/10yr$ shows that, as expected, the average recovery is still biased to higher $\gamma_{\rm GWB}$ values. In contrast, the $A_{\rm GWB}$ recovery is now less affected by the lack of power at higher frequencies, resulting in a combined posterior that agrees very well with the nominal amplitude value of the simulated GWB. The difference between the {\it SMBHB\_set} and the {\it createGWB\_set} also becomes not statistically significant. We refer to Sec. 2 of \citet{EPTA_V} for further details on this point.

In this paper, we focus on testing the recovery of the GWB signal from PTA datasets. However, pulsar's intrinsic noise parameters are also free parameters in our inference runs and, thus, subject to possible biases in the recovery. We refer to Appendix A for a brief discussion on the recovery of pulsar's intrinsic RN parameters.

From the parameter estimation analysis carried out on the {\it createGWB\_set} and on the {\it SMBHB\_set}, we can build the distribution of the BFs of HD vs CURN obtained with the reweighting method in each realisation (see Sec.~\ref{sub:recovery} for more details). Results are shown in Fig.~\ref{fig:logBF_2.4}. The median of the ${\rm log}_{10}{BF}$ distribution for the {\it createGWB\_set} is $\sim 0.5$, while for the {\it SMBHB\_set} is $\sim 0.8$. The estimated BF for EPTA \texttt{DR2new} is $\sim 60$ and is represented in the plot by the black vertical line. We note that this value is included in the 16th-84th percentile interval of the distribution for the {\it SMBHB\_set}.

Finally, we also plot the recovered angular correlation in the timing residuals, following the procedure described in the previous section. In Fig.~\ref{fig:HD_createSMBHB} we compare the mean and variance of the reconstructed correlation function to the HD curve and its cosmic variance \citep{Allen_2023}. In both cases, not only the different points are always compatible with the predicted HD correlation, but also the computed mean in each angular bin is typically included within the predicted cosmic variance.

\section{A close look at astrophysical variance: Notable realisations of the {\it SMBHB\_set}
}\label{s:specialcases}

So far, we focussed on the collective properties of the recovered signal, highlighting possible systematic biases due to the idealised model used in the analysis pipeline. It is also interesting, however, to have a close look at the 'zoology' of signals arising from a realistic SMBHB population, to get an idea of how specific features are reflected in the outcome of the analysis. We focus here on four selected realisations of the {\it SMBHB\_set}, stressing that all of them are independent statistical realisations of the same underlying astrophysical model.

\subsection{{\it Case1} and {\it Case2}: Different spectra resulting in similar posteriors}

\begin{figure*}
    \centering
	\includegraphics[width=0.57\textwidth]{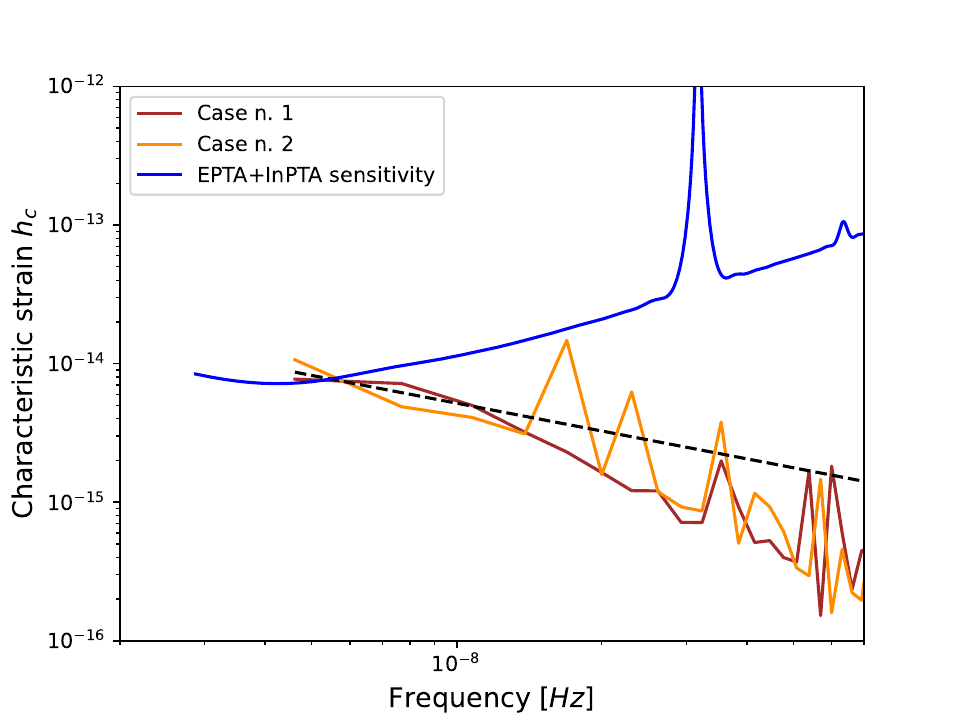}
	\includegraphics[width=0.4\textwidth]{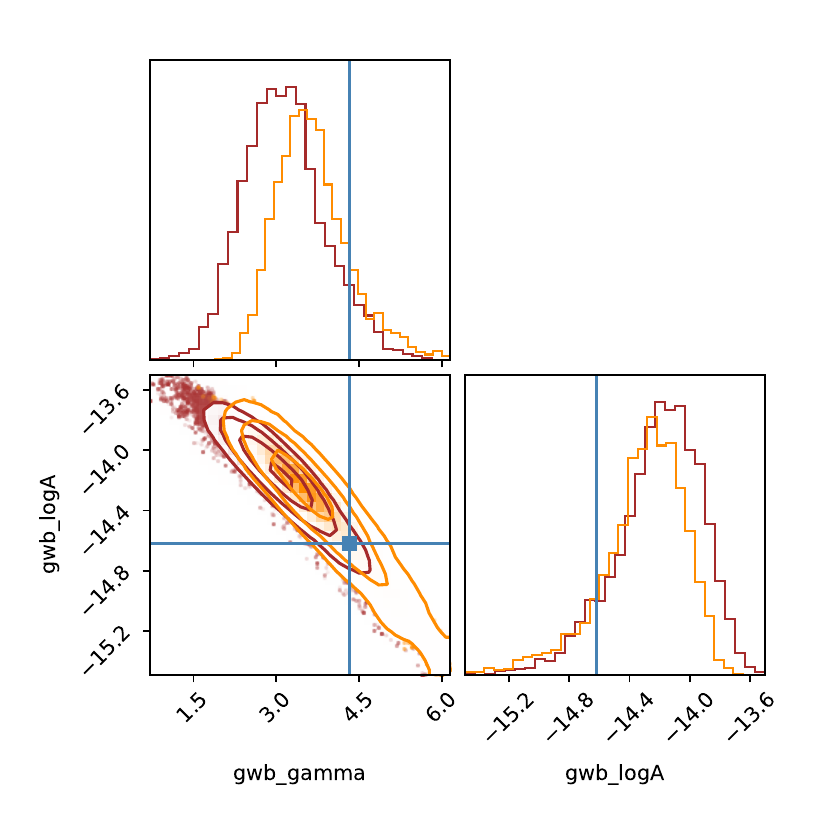}
    \caption{Comparison between the two realisations denoted as {\it Case1} and {\it Case2}. In the left panel we compare the characteristic spectra with the the $f^{-2/3}$ trend, represented by the dashed black line. The sensitivity curve is the one presented in \citet{3Ppaper}. Right panel: corner plot of the GWB parameter posteriors for {\it Case1} (brown posterior) and {\it 2} (orange posterior). The simulated GWB amplitude is $2.4\times10^{-15}$.}
    \label{fig:hc_1e2}
\end{figure*}

The first two selected cases are characterised by the spectra shown in the left panel of Fig.~\ref{fig:hc_1e2}. While {\it Case1} (brown line) has a flat, smooth bump in the lowest frequency bins, {\it Case2} (orange line) is characterised by a jagged behaviour due to few loud sources. In the right panel of Fig.~\ref{fig:hc_1e2} we show the GWB parameter posteriors obtained for those two realisations. The recovery model is the one described in Sec.~\ref{sub:recovery}. Although the two spectra are very different, the inference runs produce very similar posteriors for the GWB amplitude and slope. 
This can be qualitatively understood by comparing the spectra to the sensitivity curve associated to the considered PTA (blue curve in Fig.~\ref{fig:hc_1e2}, left panel).
In {\it Case1}, the array is mostly sensitive to the last couple of frequency bins, where the signal is flatter than the nominal $f^{-2/3}$ power law. This leads to a posterior with a slight preference for small $\gamma$ and high $A_{\rm GWB}$. Conversely, in {\it Case2}, besides detecting the signal at the lowest frequency bin, the inference pipeline picks some correlated power due to the marginally detectable loud source at $f\approx 1.5\times 10^{-8}$Hz. Since the built-in model is a single power law, this extra high-frequency power also results in a posterior with a slight preference for small $\gamma$ and high $A_{\rm GWB}$.

It is also interesting to notice that the recovered GWB parameters for these two selected realisations are quite consistent to the \texttt{DR2new} posteriors presented in Fig. 1 of \citet{EPTA_III}. This just exemplifies that there is no conflict between the signal observed in the latest PTA data and astrophysical expectations.

\subsection{{\it Case2} and {\it Case3}: Loud and louder sources.}
\label{ss:case2e3}

\begin{figure}
	\includegraphics[width=0.9\columnwidth]{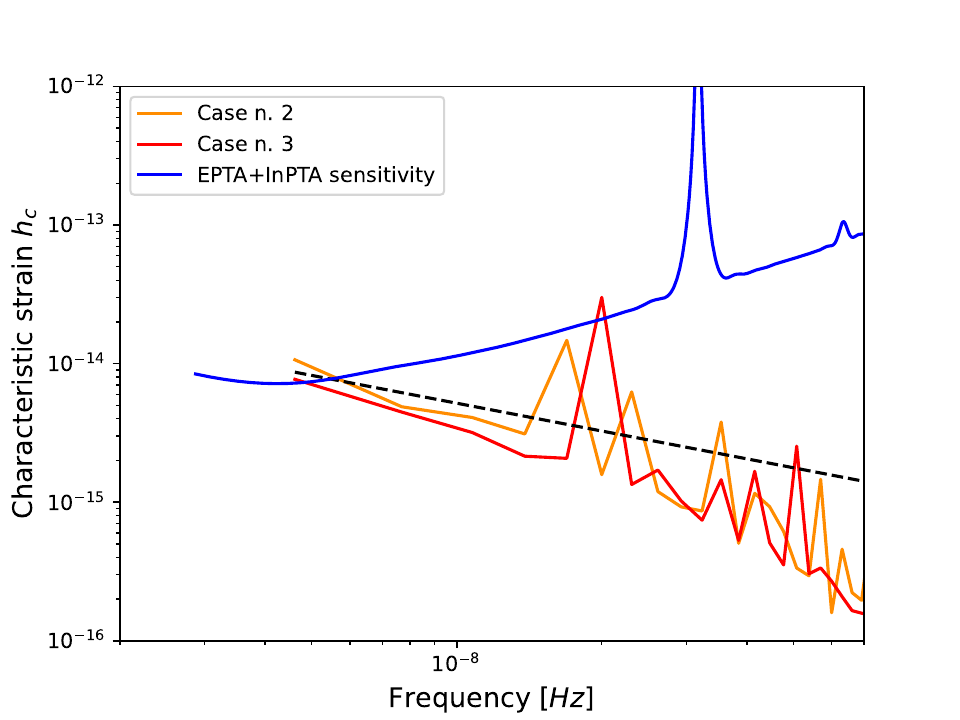}
        \vspace{-0.4cm}
	\includegraphics[width=0.9\columnwidth]{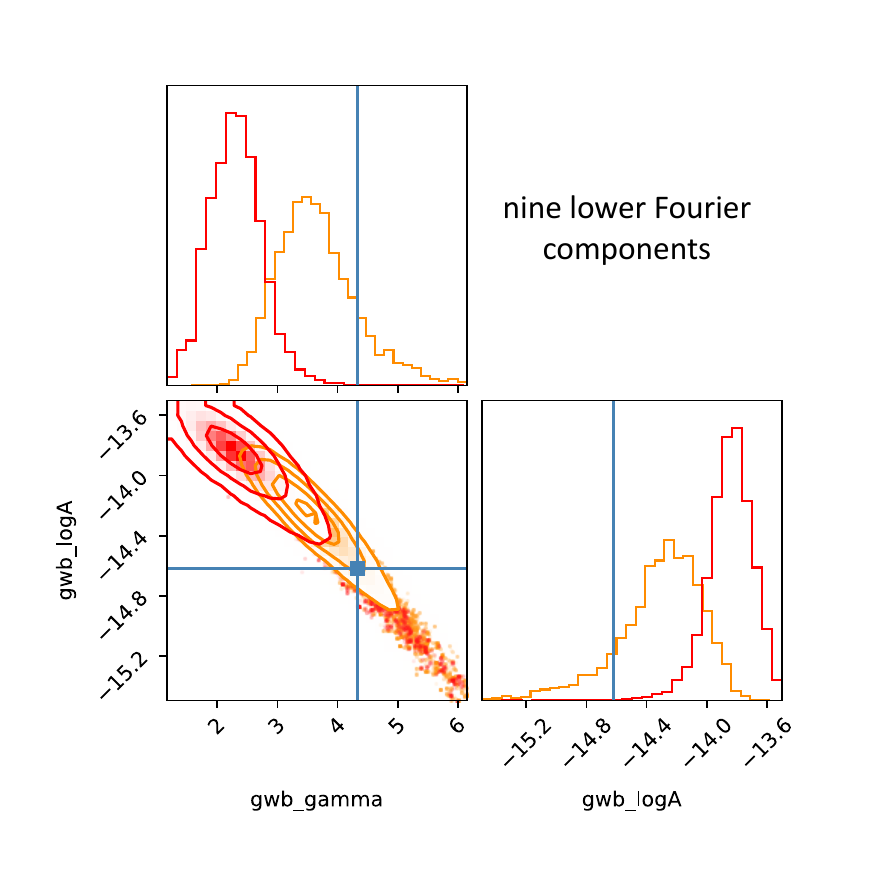}
        \vspace{-0.4cm}
	\includegraphics[width=0.9\columnwidth]{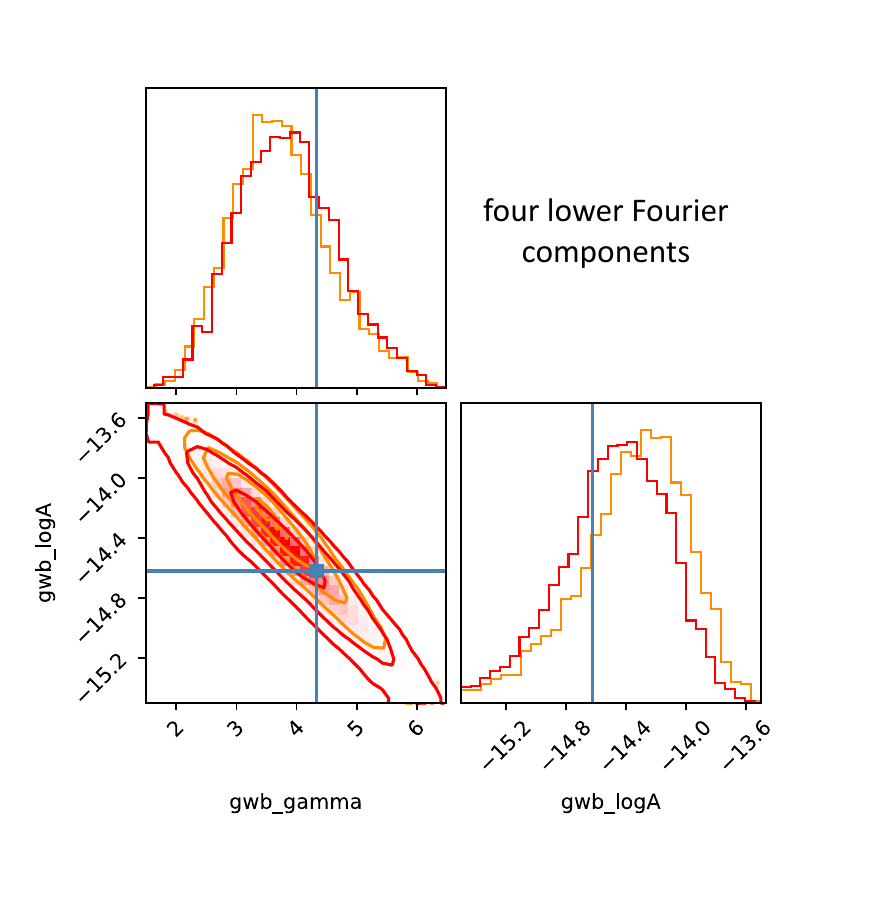}
       \caption{Notable realisations {\it Case2} (orange) and {\it Case3} (red). Top panel: GWB spectra. The dashed black line represents the $f^{-2/3}$ trend. The sensitivity curve is the one presented in \citet{3Ppaper}. The central and bottom panels show corner plots of the GWB parameter posteriors inferred from the analysis using nine (central panel) and four (bottom panel) frequency bins for the GWB recovery.}
    \label{fig:hc_2e3}
\end{figure}

To further demonstrate the impact of sparse, particularly massive and (or) nearby SMBHBs on the GWB signal recovery, we selected two realisations featuring some prominent loud sources. Those are {\it Case2}, already introduced in the previous subsection, and {\it Case3}.  The characteristic strain frequency spectra of those two realisations are presented in the top panel of Fig.~\ref{fig:hc_2e3}. The loud source at $f\approx 2\times 10^{-8}$ present in {\it Case3} produces a peak in the power that is marginally above the nominal sensitivity of the simulated PTA, we can therefore expect a strong impact on the recovered signal.

This is shown in the central and bottom panels of 
Fig.~\ref{fig:hc_2e3}. When modelling the signal with the nine lower Fourier components (frequency bins) in the GWB search, the presence of this peak biases the recovery towards low values of $\gamma_{\rm GWB}$ and high amplitudes (red posterior in the central panel). If, instead, we consider only the first four frequency bins, the bright source falls outside the frequency domain of the model and, as expected, the recovered posteriors are much closer to the nominal values for the injected SMBHB population (red posterior in the bottom panel). A similar effect, although to a lesser extent, is seen for {\it Case2}. In this case, the loud source is less prominent, contributing only marginally to the overall detected power. Still, by comparing the orange posteriors in the central and bottom panels of Fig.~\ref{fig:hc_2e3}, we can see a significant shift of the posterior to the lower right when restricting the model from nine to four frequency bins. The possible bias towards a flatter GWB spectrum due to a particularly bright GW source has been recently discussed also in~\citet{Becsy_2023}. Here we provided concrete examples of the effect on the recovery of the GWB spectra when bright sources are involved.

\subsection{{\it Case4}: A familiar HD correlation recovery}

\begin{figure*}
    \centering
	\includegraphics[width=\columnwidth]{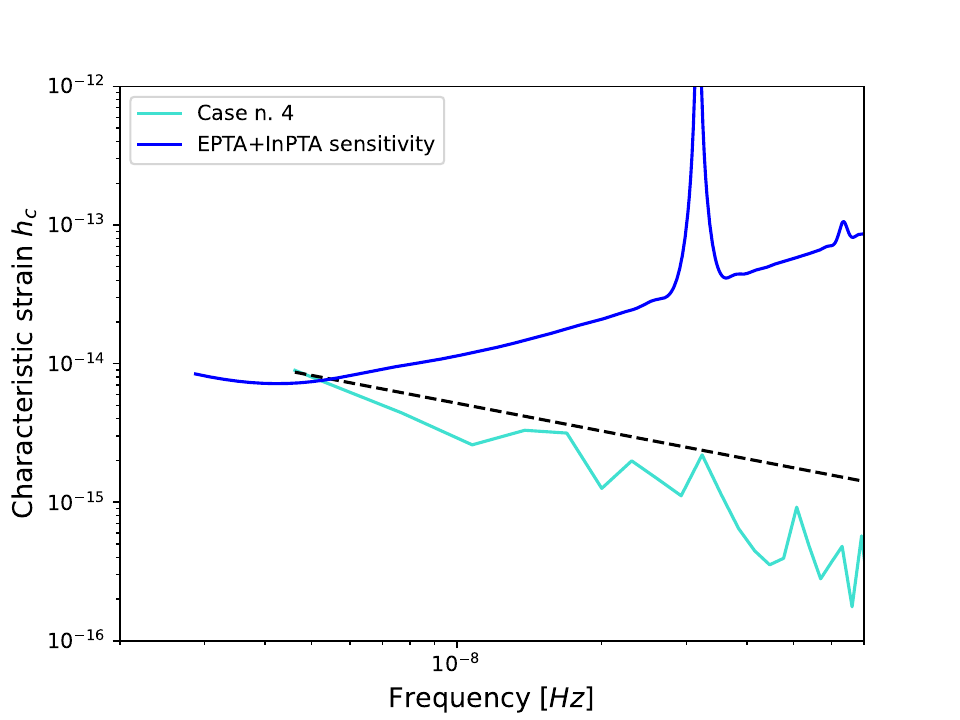}
	\includegraphics[width=\columnwidth]{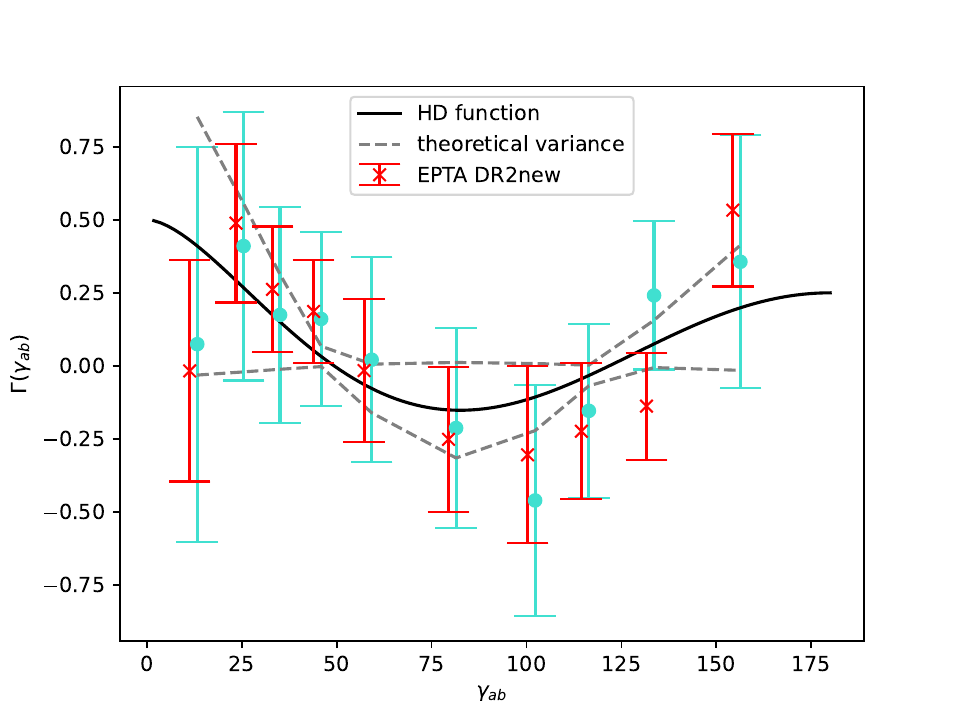}
    \caption{{\it Case4} analysis. The left panel shows the  characteristic strain spectrum compared to the $f^{-2/3}$ trend (dashed black line). The sensitivity curve is the one presented in \citet{3Ppaper}. Right panel: comparison between the recovered angular correlation in the residuals for this specific realisation (light blue points) and the results for the EPTA \texttt{DR2new} dataset (red). The expected variance of the optimal correlation estimator, for the EPTA 25 best pulsars array, is derived from prescriptions in \citet{AllenRomano}.}
    \label{fig:hc_4}
\end{figure*}

Finally, we present a fourth interesting realisation. In {\it Case4} the characteristic strain as a function of frequency does not present any particularly pronounced peak (see the left panel in Fig.~\ref{fig:hc_4}). The interesting feature observed while analysing this PTA dataset is the induced angular correlation in the residuals. In Fig.~\ref{fig:hc_4}, right panel, we show the results obtained with the OS packages (as described in Sec.~\ref{sub:recovery}) for this specific realisation (light blue points), and compare them with the results obtained for the EPTA \texttt{DR2new} dataset (red points). Each data point of the plot represents the average correlation in a bin containing 30 pulsar pairs. The averages over pulsar pairs are computed following the prescriptions in \citet{AllenRomano}. 

The reconstructed angular correlation closely follows the HD prediction. What caught our attention is the close resemblance between the right panel of Fig.~\ref{fig:hc_4} and the results from the EPTA \texttt{DR2new} dataset \citep[see Fig.~6 of ][]{EPTA_III}. In both plots, the considered pulsars are the same (EPTA best 25 pulsars) and the correlation is computed following the same procedure. 

The similarity in the reconstructed HD is also reflected in the estimated BF for HD correlated common signal vs CURN. Using the reweighting technique in the inference run, we obtain BF$\approx 62$ for {\it Case4}, which is very consistent with the BF$\approx 60$ found in EPTA \texttt{DR2new} \citep[][]{EPTA_III}. 

The BF (HD vs CURN) of those four notable cases are summarised in Table~\ref{tab:bf} for completeness, and they give a flavour of the role played by stochasticity in the estimate of signal significance and parameters. For example, the spectra of {\it Case1} and {\it Case4} look very similar, with perhaps {\it Case1} showing a bit more power in the lowest frequency bins. Still, in this case, the result of the analysis is inconclusive (BF$=1.179$ for HD vs CURN), whereas in {\it Case4} there is strong evidence of an HD correlated process (BF$=62.5$). This is because, in these early stages of detection, the output of the analysis is very sensitive to the specific realisation of the noise processes and to the specific sky locations of the loudest systems contributing to the GWB with respect to the best pulsars in the array. The key role played by the stochastic realisation of the noise processes involved is also demonstrated by the BF distributions shown in Fig.~\ref{fig:logBF_2.4}; even when we inject a nearly ideal signal with {\it createGWB}, the distribution of Bayes factors returned by the analysis spans several orders of magnitudes.

\begin{table}
\begin{center}
  \begin{tabular}{|c|c|c|c|}
  \hline
  Case n. & \, ${\rm BF}^{\rm HD}_{\rm CURN}$ \\
  \hline
  \hline
  1 & 1.179 \\
  \hline
  2 & 12.358 \\
  \hline
  3 & 70.133 \\
  \hline
  4 & 62.503 \\
  \hline
  \end{tabular}
  \caption{BF obtained from the reweighted nine-frequency bins inference runs \citep{Hourihane_2023} for the four notable cases analysed in Sec.\ref{s:specialcases}. The BF is computed for an HD-correlated signal over a common uncorrelated red noise. \label{tab:bf}}
\end{center}
\end{table}

\section{Discussion and conclusions}
\label{s:conclusion}

In this paper, we carried out an extensive investigation of the performance of current PTA GW analyses on simulated PTA datasets injected with different types of GW signals. Our simulations included realistic levels of white noise, red noise, and DM variations, that were gauged to create mock data equivalent to the recently published EPTA \texttt{DR2new}. 
We injected in those data either a stochastic, stationary, Gaussian GWB with a standard $f^{-2/3}$ power-law spectrum (the {\it createGWB\_set} of simulations) or the incoherent superposition of sinusoidal signals from a cosmic population of SMBHBs (the {\it SMBHB\_set} of simulations), paying particular attention to possible limitations and biases arising from the mismatch between the signal present in the data and the model used for the inference. The injected signals were calibrated on the results of \cite{EPTA_III}, where the amplitude of the signal was estimated to be $A{\rm GWB}\approx 2.4\times10^{-15}$ for a power-law spectrum with $\gamma_{\rm GWB}=-13/3$.

We quantified the performance of the analysis model by constructing P\textendash P plots for the amplitude of the recovered signal for each set of simulations. The {\it createGWB\_set} demonstrates that, when the model matches with the injected signal, the outcome of the analysis is reliable, although some mild bias can arise due to the complex multi-dimensional nature of the parameter space that needs to be searched over. In fact, when fixing all the parameters but the GWB amplitude, the estimate of this parameter is unbiased; conversely, when performing a joint search on the GWB and noise parameters (including RN and DM) the signal amplitude is slightly biased towards lower amplitudes and steeper spectra. Such bias has also been seen in simulations of individual pulsar noise analysis \citep{EPTA_II}, and although it is still between the one- and two-$\sigma$ level (Fig~\ref{fig:mainPP}, \ref{fig:combinedAG}), it requires further investigation. 

In the case of the more realistic {\it SMBHB\_set}, where there is a mismatch between the injected signal and the simplified recovery model, the bias is much more prominent and the recovered spectra for the common noise tend to be steeper (higher $\gamma_{\rm GWB}$) and with lower amplitude than the injected signal (Fig~\ref{fig:mainPP}, \ref{fig:combinedAG}). 
This is due to the nature of the SMBHB population, which features a large number of weak sources with a tail of sparse, loud systems, as discussed in Sec.~\ref{ss:createvsrealinj}. When one of these loud systems is present in the SMBHB population, the recovered GWB spectra can appear flatter (lower $\gamma_{\rm GWB}$) and with a higher amplitude. 

Using the reweighting method, we were also able to build a distribution of BF from the different realisations of both the {\it createGWB\_set} and the {\it SMBHB\_set} (see Fig.~\ref{fig:logBF_2.4}). We found no significant difference between the two distributions, although the {\it SMBHB\_set} results on average in slightly higher BFs. We note that in both sets, the ${\rm log}BF$ distribution has a large scatter, with one-$\sigma$ confidence interval spanning index covering the $[0,2]$ interval. We also computed the induced angular correlation in the timing residuals of the different datasets and showed that they agree, within the predicted variance, with the HD correlation.

These results allow us to make several interesting considerations. First, as already shown in \cite{Cornish_2016} the mismatch between the model and the data does not seem to affect our ability to recover the GW signal. This is probably because the detection significance is based on the intra-pulsar correlation properties of the signal (i.e. the HD overlap reduction function) which is a feature that emerges for any collection of GW signal, regardless on its specific properties (e.g. stationarity, Gaussianity, isotropy, spectral shape). Conversely, the reconstruction and interpretation of the observed signal can be severely biased by the use of a simplified GWB model. For example, while loud individual sources can cause a flattening of the spectrum which might erroneously misinterpreted as environmental effects or high eccentricity, the lack of them might result in a steep inferred spectrum which can be (again erroneously) claimed to be inconsistent with an astrophysical origin. Finally, the stochastic nature of the noise has a major impact on the outcome of the analysis. Even for the {\it createGWB\_set}, where we effectively always inject the same signal, the analysis can either return a detection supported by strong evidence or an inconclusive result, just depending on different realisation of the stochastic process describing the noise.

Within this diverse and complex phenomenology, we highlighted also some notable realisations that exemplifies some of the possible analysis outcomes. In particular, we highlighted that: (i) little astrophysical information can be drawn from an inference run using a simple power-law GWB model, by showing that very different GW signals can result in similar inferred GWB parameters, (ii) the presence of loud sources can introduce a bias in the recovery of the common process, (iii) some realisations of our simulations result in a recovered HD correlation and BF completely in line with what observed in EPTA DR2.

Finally, \cite{Becsy_2023} presented a similar set of simulations based on the NANOGrav 15yr dataset. They also use astrophysically motivated SMBHB populations to generate the GW signal and carry out a thorough analysis of a realistic  dataset including unevenly sampled data and pulsar red noise. They perform Bayesian inference from the data, computing HD vs CURN Bayes factors and conclude that the simple GWB model implemented in the current analysis is able to recover a realistic GW signal, although they stress that loud sources might affect the inference. Compared to their work, our investigation adds several layers of sophistication. Our simulated data also include observations at two frequencies per epoch, allowing the inclusion of DM as a further source of noise, which allows as to test the analysis performance on a more complicated situation, closer to the real data. We cast our results in terms of P\textendash P plots, and we compute combined posteriors of several realisations of the same dataset, which allowed us to identify some interesting systematic biases in the recovered signal. Finally, we carried out a systematic comparison on the analysis performed on realistic signal injections vs an ideal GWB generated by the {\it createGWB} function, which allowed us to identify potential difficulties due to the signal vs template mismatch.  

Now that evidence of a GW signal is  emerging independently from several PTA data, it is important to 
assess
the reliability of our analysis methods, in order to maximise the astrophysical potential of the PTA experiments. The present work, along with  \cite{Becsy_2023}, represents an important first step in this direction, which needs to be extended to include increasingly realistic situations. For example, in this article, we did not investigate complex signal spectra due to environmental effects or binary eccentricity. Likewise, we used the exact same model for noise injection and recovery. As shown in \cite{EPTA_V}, the excess or mis-modelled noise in the data can be absorbed within the common correlated signal in the analysis, leading to further biases and interpretation issues. Finally, including single sources along with a GWB in the recovery model might significantly improve the quality of the inference, especially when prominent peaks are present in the GW spectrum. \\

The main scripts used to analyse the data sets here described can be found at \texttt{https://github.com/ serevaltolina/PTAsim\_fromSMBHBpop}. The {\it SMBHB\_set} is also available on \texttt{zenodo} under the following Digital Object Identifier:  \texttt{10.5281/zenodo.10276364}. In particular, for each universe realisation we uploaded: the SMBHBs specifications, the chain file obtained from the inference runs (while sampling for each pulsar intrinsic noise parameters and a common uncorrelated red noise), each pulsar .par and .tim files.

\section*{Acknowledgements}
The authors acknowledge the support of colleagues in the EPTA. We thank Jonathan Gair and Lorenzo Speri for their insightful discussions and feedback since the very beginning of this project. We are grateful to Siyuan Chen for providing us with the maximum likelihood estimates of the individual pulsar RN and DM noise models when searched for along with a common process. We also thank Bruce Allen and Rutger Van Haasteren for useful conversations.
This work was supported by the Max-Planck-Gesellschaft (MPG) and the ATLAS cluster computing team at AEI Hannover.
AS and GS acknowledge the financial support provided under the European Union’s H2020 ERC Consolidator Grant ``Binary Massive Black Hole Astrophysics" (B Massive, Grant Agreement: 818691). ASa thanks the Alexander von Humboldt Foundation in Germany for a Humboldt fellowship for postdoctoral researchers. 

We also made use of \texttt{numpy} \citep{Harris_2020}, \texttt{matplotlib} \citep{matplotlib}, \texttt{corner} plot \citep{corner}, \texttt{scipy} \citep{scipy}, \texttt{tempo2} \citep{Hobbs_2006,Edwards_2006}, \texttt{libstempo} \citep{libstempo}, \texttt{enterprise} \citep{Ellis_2019} and the \emph{Optimal Statistic} framework \citep{Anholm_2009,Demorest_2013,Chamberlin_2015,Vigeland_2018}.

\bibliographystyle{aa}
\bibliography{manuscript}

\begin{appendix}
\section{Intrinsic noise parameters recovery}

\begin{figure}
	\includegraphics[width=\columnwidth]{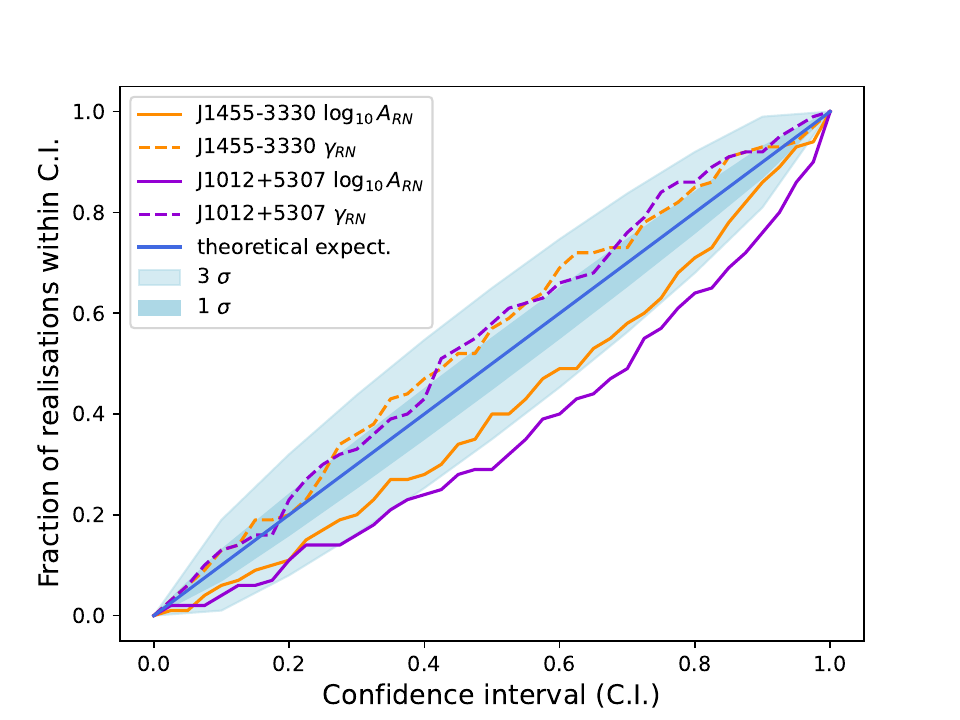}
    \includegraphics[width=\columnwidth]{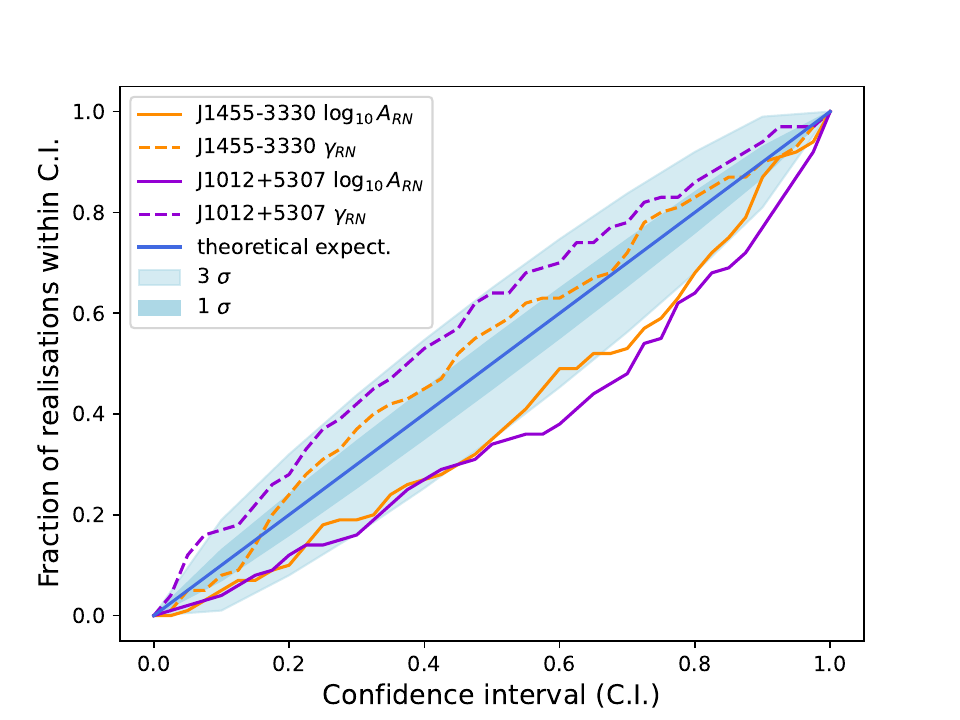}
    \caption{P--P plots for the recovery of pulsars J1455-3330 and J1012+5307 RN parameters from the MCMC inference runs performed on the {\it createGWB} (top panel) and {\it SMBHB\_set} (bottom). Dashed lines are for the power-law slope, while the solid lines refer to the RN amplitude. }
    \label{fig:PPnoise}
\end{figure}

While we focussed on testing the viability and robustness of the recovery of the GWB signal, it is also interesting to briefly look at the recovered values for the pulsar's intrinsic RN when a common signal is included in the model. In particular, we choose the pulsars J1455-3330 and J1012+5307 (see Table~\ref{tab:psrsnoise} for the injected noise). We used the posteriors obtained from the MCMC runs over the {\it createGWB} and the {\it SMBHB\_set} (see Sec.~\ref{ss:createvsrealinj}) to build the P\textendash P plots for the inferred RN amplitude and slope for those two pulsars. The results are shown in Fig.~\ref{fig:PPnoise}.

In this case there is no obvious distinction between the results from the {\it createGWB} and the {\it SMBHB\_set}. In fact, the intrinsic pulsars RN is modelled in the same way in both of them and the different GWB injected does not seem to affect the RN inference. However, in both cases, the recovered distributions are slightly biased. Even if they all lie in a range comparable with the three-$\sigma$ confidence interval, the distributions for ${\rm log}A_{\rm RN}$ are constrained below the diagonal, while the ones for $\gamma_{\rm RN}$ are always above. This means that, for both pulsars, on average the recovered RN has spectra biased towards lower amplitudes and higher slopes; the same type of bias that we obtained for the GWB recovery.

\citet{EPTA_II} carried out a similar test on the RN recovery. In their Fig.~6, they show a P\textendash P plot test on their noise parameter estimation from simulated data containing pulsar intrinsic noise only. Interestingly, they observe a bias in the \emph{opposite direction} with respect to our results. Their conclusion is that, when the spectral slope of RN is greater than $\sim 4$, the recovery is, on average, biased towards flatter spectra with higher amplitudes. In our case, the spectral slope of the tested pulsars was much lower than that threshold (about 1.4 for both pulsars, see Table~\ref{tab:psrsnoise}). This may be the at the root of the different results.
We defer a deeper investigation of those mild biases to future  studies.

\end{appendix}

\end{document}